\documentclass[runningheads]{llncs}
\usepackage[caption = false]{subfig}
\usepackage{graphicx}
\usepackage{textcomp}
\usepackage{multicol}
\usepackage{algorithm}
\usepackage{algorithmic}
\usepackage[english]{babel}
\usepackage{url}
\usepackage{multirow}
\begin{document}
\title{Event Detection and Retrieval on Social Media}

\author{Manos Schinas\inst{1,2} \and Symeon Papadopoulos\inst{1} \and
Yiannis Kompatsiaris\inst{1} \and Pericles Mitkas\inst{2}}

\institute{
	Centre for Research and Technology Hellas \\
	\email{\{manosetro,papadop,ikom\}@iti.gr}\\
\and
	Aristotle University of Thessaloniki \\
	\email{manosetro@issel.ee.auth.gr, mitkas@eng.auth.gr}
}

\authorrunning{M. Schinas et al.}

\maketitle

\section{Introduction}
\label{sec:intro}

\begin{figure}
\centering
\includegraphics[width=0.8\textwidth]{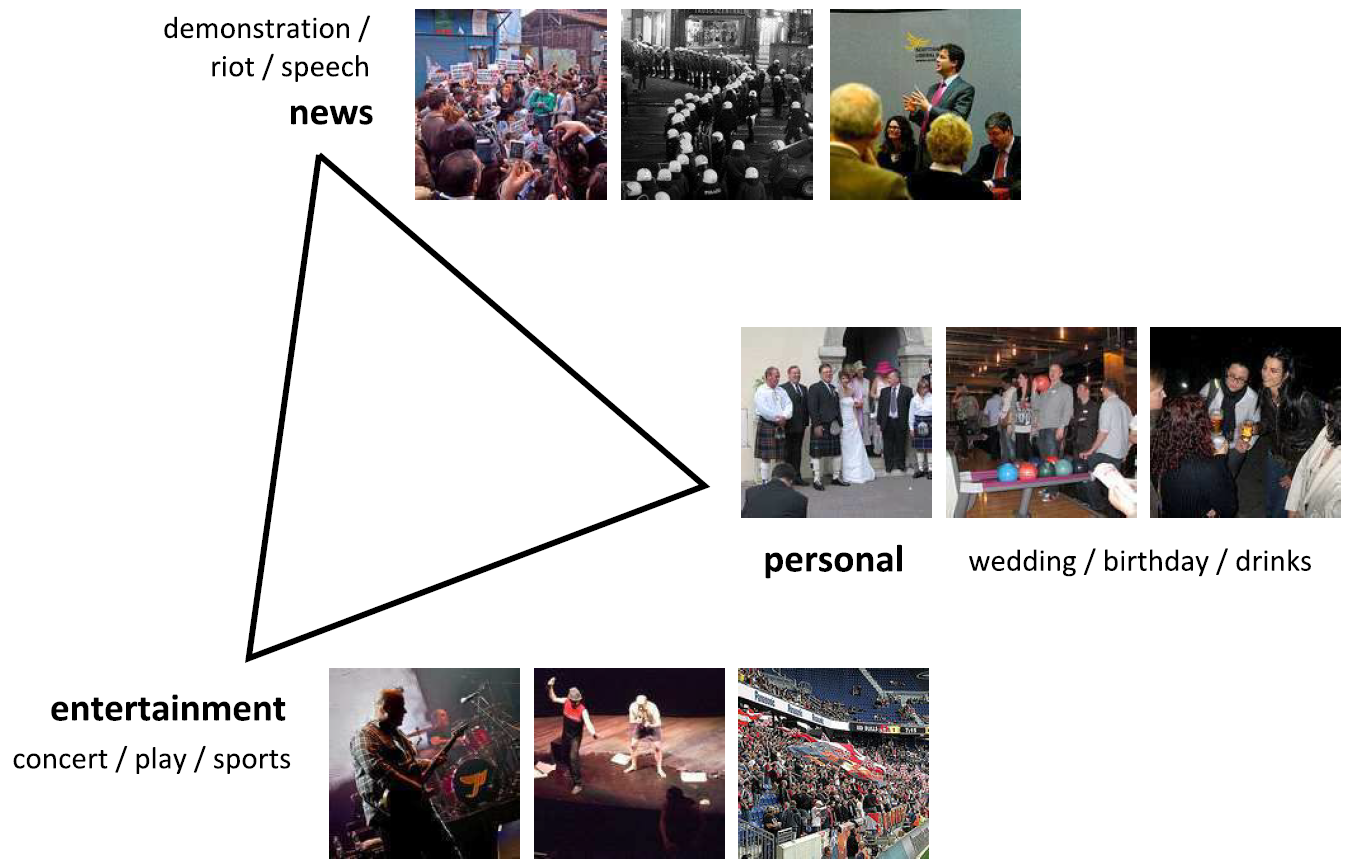}
\caption{Examples of event categories and images. Figure originally appeared in \cite{petkos_sewm2014}.}
\label{fig:event_examples}
\end{figure}

In the recent years, we have witnessed the rapid adoption of social media platforms, such as Twitter, Facebook and YouTube, and their use as part of the everyday life of billions of people worldwide. Given the habit of people to use these platforms to share thoughts, daily activities and experiences it is not surprising that the amount of user generated content has reached unprecedented levels, with a substantial part of that content being related to real-world \textit{events}, i.e. actions or occurrences taking place at a certain time and location. 
%
Figure \ref{fig:event_examples} 
illustrates three main categories of events along with characteristic photos from Flickr\footnote{\url{https://www.flickr.com/}} for each of them: a) news-related events, e.g. demonstrations, riots, public speeches, natural disasters, terrorist attacks, b) entertainment events, e.g. sports, music, live shows, exhibitions, festivals, and c) personal events, e.g. wedding, birthday, graduation ceremonies, vacations, and going out. 

Depending on the event, different types of multimedia and social media platform are more popular. For instance, news-related events are extensively published in the form of text updates, images and videos on Twitter and YouTube, entertainment and social events are often captured in the form of images and videos and shared on Flickr and YouTube, while personal events are mostly represented by images that are shared on Facebook and Instagram.

Given the key role of events in our life, the task of annotating and organizing social media content around them is of crucial importance for ensuring real-time and future access to multimedia content about an event of interest. However, the vast amount of noisy and non-informative social media posts, in conjunction with their large scale, makes that task very challenging. 
For instance, in the case of popular events that are covered live on Twitter, there are often millions of posts referring to a single event, as in the case of the World Cup Final 2014 between Brazil and Germany, which produced approximately 32.1 million tweets with a rate of 618,725 tweets per minute\footnote{\url{https://www.engadget.com/2014/07/14/brazil-world-cup-final-records/}}. Processing, aggregating and selecting the most informative, entertaining and representative tweets among such a large dataset is a very challenging multimedia retrieval problem. In other cases, the challenge is to discover sets of multimedia items that refer to the same event within a much larger dataset. That is, for instance, the case with the Yahoo-Flickr Event Summarization Challenge\footnote{\url{http://www.acmmm.org/2015/call-for-contributions/multimedia-grand-challenges/}}, which called for methods that automatically discover events in the YFCC100M collection \cite{thomee2016yfcc100m}, which contains
approximately 99 million photos and 1 million videos. In addition to the large scale of the problem, a challenge in trying to detect events in this collection stems from the inconsistent quality of user provided metadata: for instance, several images have no title or description, are erroneously tagged and are missing geographical information.

In this chapter, we present several research efforts from recent years that tackle the two main problems that were discussed above: 
\begin{itemize}
\item \textbf{Event detection.} Given archived collections or live streams of social media items, the purpose of these methods is to identify previously unknown events in the form of sets of items that describe them. In general, the events could be of any type, but there are also approaches aiming at events of specific type. 
\item \textbf{Event-based media retrieval and summarization.} Given a target event the goal is first to identify relevant content and then to represent it in a concise way, selecting the most appealing and representative content. 
\end{itemize}
In the following, we will always be referring to events $E$ in relation to \textit{collections} $\mathcal{C}$ or \textit{streams} $\mathcal{S}$ of multimedia items (images, videos). Individual multimedia items will be denoted as $c$ (standing for \textit{content}), and they will refer to tuples of metadata fields, e.g. a given content item will be denoted as $c = \{u,  t, g, l, d, T, v\}$, where $u$ denotes the social media user who posted the particular content, $t$ the publication time, $g$ its geographical location (latitude, longitude), if available, $l$ its title (label), $d$ its description, $T$ a set of associated tags, and $v$ its visual content (image, video). The particular fields have been defined based on Flickr, since Flickr is among the most common multimedia sharing platforms, where event detection research is performed. However, very similar representations are applicable to other platforms, e.g. in the case of Twitter, we have a very similar representation, with the difference that $l$ refers to the tweet text, there is no description $d$, $T$ refers to the set of accompanying hashtags, and $v$ refers to the embedded media content. When necessary the above notation will be used with subscripts to differentiate between different instances, e.g. event $x$ will be denoted as $E_x$ and media item $i$ as $c_i$. For our discussion, an event corresponds to a set of multimedia items that depict scenes from the event, i.e. $E = \{c\}$.

The remainder of this chapter is structured as follows: Section \ref{sec:event_detection} reviews the rich literature on the topic of event detection, organizing methods into feature-pivot (subsection \ref{sec:feature_pivot}), document-pivot (subsection \ref{sec:doc_pivot}), and topic modeling ones (subsection \ref{sec:topic_models}). Then, Section \ref{sec:retrieval_summarization_methods} discusses a number of methods that tackle the problems of event-oriented retrieval (subsection \ref{sec:event_based_retrieval}) and summarization (subsection \ref{sec:event_based_summarization}). Section \ref{sec:datasets_evaluation} provides an overview of evaluation measures and datasets that are used for evaluating event detection and retrieval methods, along with a comparative view of reported results. Finally, the chapter concludes in Section \ref{sec:conclusions}.


\section{Event Detection}
\label{sec:event_detection}

Event detection is the problem of associating multimedia items (images, videos) in a collection $\mathcal{C}$ or stream $\mathcal{S}$ to a real-world event $E$. Typically, the event is not known in advance and is the result of the event detection process. There is a large variety of methods for event detection, which we group into three categories based on their \textit{principle}: 
\begin{itemize}
\item \textit{Feature-pivot:} These are based on detecting abnormal patterns in the appearance of features, such as words. Once an event happens, the expected frequency of a feature will be abnormal comparing to its historical behavior, indicating a potential new event.
\item \textit{Document-pivot:} This category comprises methods that represent documents as items to be clustered using some similarity measure. 
\item \textit{Topic modeling:} This includes methods that utilize statistical models to identify events as latent variables in the documents.  
\end{itemize}
A further attribute of methods pertains to their applicability on static collections versus streams of documents (\textit{static vs stream}). Based on this, the methods can be divided into two main categories: \textit{Retrospective Event Detection} (RED) and \textit{First Story Detection} (FSD). RED methods aim to detect unknown events from an accumulated set of documents, e.g. a collection of Flickr images or news articles. On the other hand, FSD methods, also known as new or online event detection, focus on the real-time discovery of events from a live stream of social media messages, e.g. tweets. There is also a third category of methods that we call timeslot-based (TSB): those are positioned between RED and FSD methods and detect events in a pseudo-incremental way.

\begin{figure}
\centering
\includegraphics[width=0.8\textwidth]{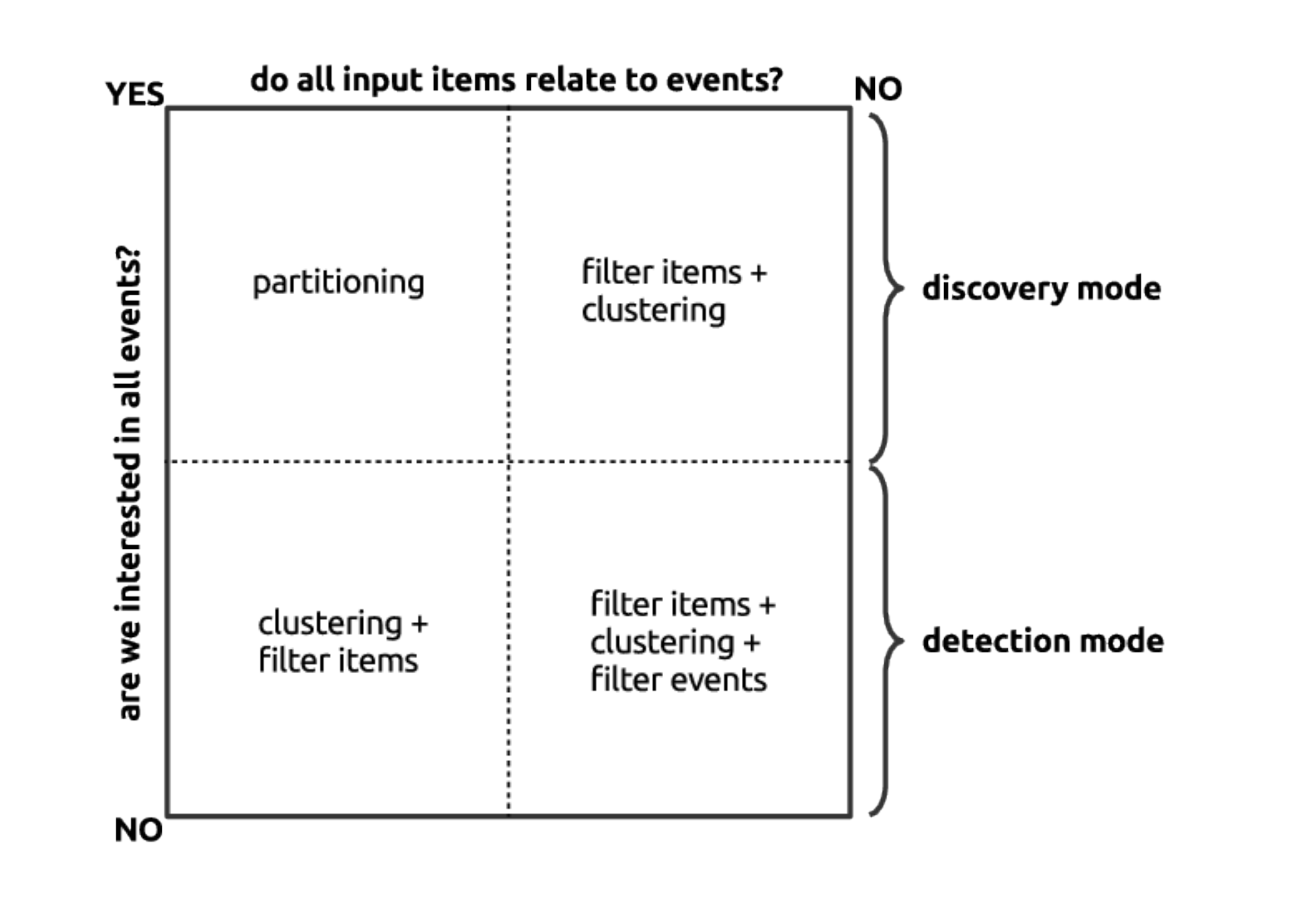}
\caption{Main event detection processing paradigms in relation to whether all items in a collection are associated with events and whether all events are of interest.}
\label{fig:event_detection_quadrant}
\end{figure}

An additional attribute for grouping methods is the use of \textit{modalities} by each method. Most methods are based on text analysis, but several methods have been proposed that integrate other modalities available in social media such as location, time, visual content, etc. 
Finally, the presented methods are also categorized by their \textit{mode}, namely \textit{discovery} versus \textit{detection} methods. Figure \ref{fig:event_detection_quadrant} illustrates a few important processing paradigms (partitioning, clustering, filtering) in relation to whether all items in a collection $\mathcal{C}$ need to be assigned to events, and whether all represented events are of interest (discovery mode) or just a subset of them (detection mode).   
Table \ref{tab:event_detection_categorization} presents an overview of the methods to be discussed in this section along with their attributes (principle, static vs stream, modalities, mode). 

\begin{table}
\label{tab:event_detection_categorization}
\caption{Categorization of event detection methods according to different attributes. The following abbreviations have been used: FEAT/DOC/TOPIC stand for feature-pivot, doc-pivot and topic modeling approaches, RED/FSD/TSB stand for Retrospective Event Detection, First Story Detection and Timeslot-based approaches, TXT/VIS/TM/US /SOC/LOC stand for Text, Visual, Time, User, Social links and Location modalities, and DETECT/DISCOV stand for Detection and Discovery mode.}

\scalebox{0.85}{
\footnotesize
\begin{tabular}{|l|l|l|l|l|} \hline
\textbf{Method} & \textbf{Principle} & \textbf{Static/Stream} & \textbf{Modalities} & \textbf{Mode} \\ \hline
Fung et al., 2005 \cite{fung2005parameter} & FEAT & TSB & TXT & DISCOV \\ \hline
He et al., 2007 \cite{he2007analyzing} & FEAT & RED & TXT & DISCOV \\ \hline
Mathioudakis \& Koudas, 2010 \cite{mathioudakis2010twittermonitor} &  FEAT & TSB & TXT & DISCOV \\ \hline
Sakaki et al., 2010 \cite{sakaki_www2010} & FEAT & TSB & TXT & DETECT \\ \hline
Weng \& Lee, 2011 \cite{weng2011event} & FEAT & TSB & TXT & DISCOV \\ \hline
Li et al., 2012 \cite{li2012twevent} & FEAT & TSB & TXT & DISCOV \\ \hline
Alvanaki et al., 2012 \cite{alvanaki2012see} & FEAT & TSB & TXT & DISCOV \\ \hline
Cataldi et al., 2010 \cite{cataldi2010emerging} & FEAT & TSB & TXT & DISCOV \\ \hline
Parikh \& Karlapalem, 2013 \cite{parikh2013events} & FEAT & RED & TXT & DISCOV \\ \hline
Chen \& Roy, 2009 \cite{chen2009event} & FEAT & RED & TXT & DISCOV \\ \hline
Sayyadi et al., 2009 \cite{sayyadi2009event} & FEAT & TSB & TXT & DISCOV \\ \hline
Guille \& Favre, 2014 \cite{guille2014mention} & FEAT & TSB & TXT & DISCOV \\ \hline
Zhang et al., 2015 \cite{zhang2015event} & FEAT & TSB & TXT & DISCOV \\ \hline
Sankaranarayanan et al., 2009 \cite{sankaranarayanan2009twitterstand} & DOC & FSD & TXT, TM & DISCOV \\ \hline
Petrovi\'{c} et al., 2010 \cite{petrovic2010streaming}  & DOC & FSD & TXT & DISCOV \\ \hline
Becker et al., 2011 \cite{becker2011beyond} & DOC & FSD & TXT & DISCOV \\ \hline
Lee, 2012 \cite{lee2012mining} & DOC & RED & TXT, TM & DISCOV \\ \hline
Petrovi\'{c} et al., 2012 \cite{petrovic2012using}  & DOC & FSD & TXT & DISCOV \\ \hline
Moran et al., 2016 \cite{moran2016enhancing}  & DOC & FSD & TXT & DISCOV \\ \hline
Aggarwal \& Subbian, 2012 \cite{aggarwal2012event} & DOC & FSD & TXT, SOC & DISCOV \\ \hline
Becker et al., 2009 \cite{becker2009event}  & DOC & RED & TXT, TM, LOC & DISCOV \\ \hline
Becker et al., 2010 \cite{becker2010learning}  & DOC & RED, FSD & TXT, TM, US, LOC & DISCOV \\ \hline
Reuter \& Cimiano, 2012 \cite{reuter2012event}  & DOC & FSD & TXT, TM, LOC & DISCOV \\ \hline
Petkos et al., 2012 \cite{petkos2012social}  & DOC & RED & TXT, VIS, TM, US, LOC & DISCOV \\ \hline
Bao et al., 2013 \cite{bao2013social}  & DOC & RED & TXT, VIS, TM, LOC & DISCOV \\ \hline
Wang et al., 2012 \cite{wang2012social}  & DOC & RED & TXT, TM, LOC 
& DISCOV \\ \hline
Petkos et al., 2017 \cite{petkos_mtap2017}  & DOC & RED & TXT, VIS, TM, US, LOC 
& DISCOV \\ \hline
Benson et al., 2011 \cite{benson2011event} & TOPIC & RED & TXT & DETECT \\ \hline
Ritter et al., 2012 \cite{ritter2012open}  & TOPIC & RED & TXT & DISCOV \\ \hline
You et al., 2013 \cite{you2013geam}  & TOPIC & RED & TXT, TM, LOC & DISCOV \\ \hline
Zhou \& Chen, 2014 \cite{zhou2014event}  & TOPIC & RED & TXT, TM, LOC & DISCOV \\ \hline
Zhou et al., 2015 \cite{zhou2015unsupervised}  & TOPIC & RED & TXT, TM, LOC & DISCOV \\ \hline
Cai et al., 2015 \cite{cai2015popular}  & TOPIC & RED & TXT, VIS, TM, LOC & DISCOV \\ \hline
Diao \& Jiang, 2013 \cite{diao2013unified}  & TOPIC & RED & TXT, US & DISCOV \\ \hline
Wei et al., 2015 \cite{wei2015bayesian} & TOPIC & RED & TXT, TM, LOC & DISCOV \\ \hline
Hu et al., 2012\cite{hu2012lda} & TOPIC & RED & TXT & DISCOV \\ \hline
\end{tabular}
}
\end{table}


\subsection{Feature-pivot methods}
\label{sec:feature_pivot}

\begin{figure}
\centering
\includegraphics[width=1.05\textwidth, trim={0 3cm 0 1cm},clip]{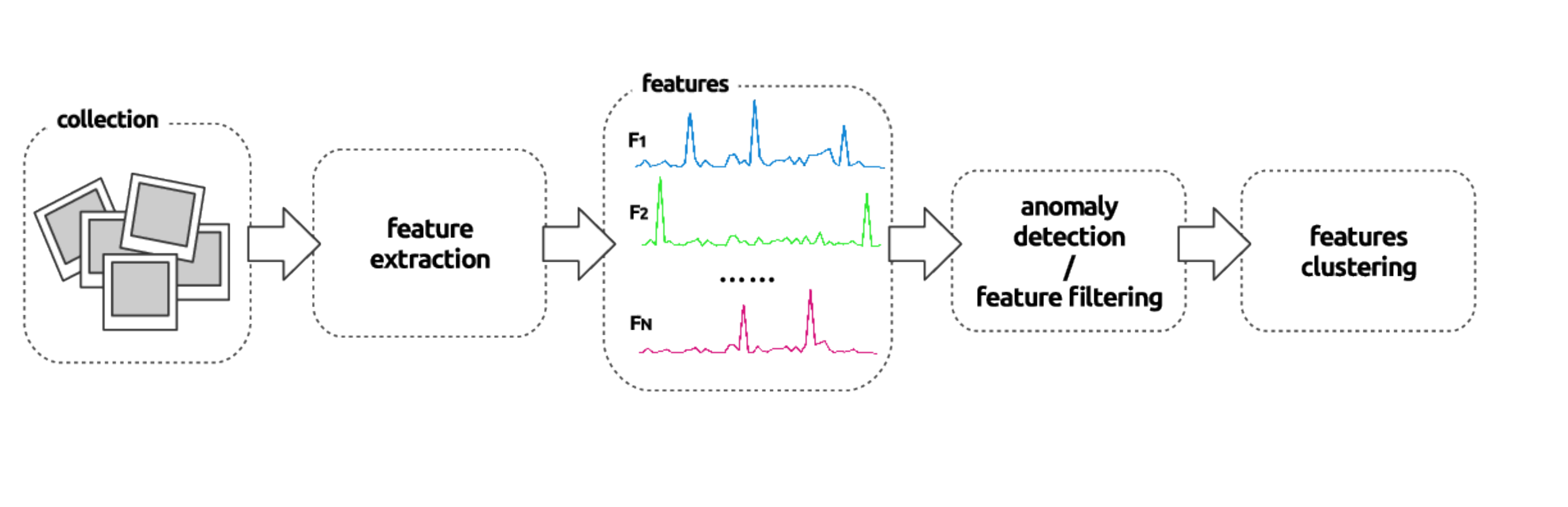}
\caption{Feature-pivot paradigm for event detection.}
\label{fig:feature_pivot_detection}
\end{figure}

This section discusses those event detection methods that rely on the detection of text features, which are likely to refer to events. These feature-pivot techniques, proposed originally for the analysis of timestamped document streams, consider an event as a bursty activity that makes some of the text features more prominent. The intuition behind these methods is that once an event emerges, certain features related to it will exhibit a similar abnormal rise in their frequency. The type of feature can range from single keywords, named entities and phrases to social interactions. Traditionally, the distributions of the extracted features over time is analyzed, anomalies in these distributions are detected, and finally, the discovery of events is conducted by grouping features that exhibit a similar behavior. This process is depicted in Figure \ref{fig:feature_pivot_detection}. According to this, an event is represented as a number of features showing an abnormality in appearance counts. The initial documents are assigned to the detected events, based mainly on the appearance of the event-related features in them. As these methods rely on the modeling of feature appearance in time, a large amount of documents must be available. Despite being applicable to collections of timestamped documents, these techniques cannot work in a pure real-time fashion as there is a need for knowledge of the behavior of feature frequencies over a certain period of time. To overcome that limitation, an incremental approach is usually followed, by detecting events on predefined timeslots. Another drawback of feature-pivot approaches is that as they depend on the detection of a bursty activity, they typically capture trends. Thus events that are not trendy and do no attract a lot of attention are likely to be missed. 

A seminal feature-pivot method was presented by Fung et al. \cite{fung2005parameter}. This aimed at detecting bursty events in text streams based on the temporal distributions of features. Although the approach is related to plain document streams (not social media), the underlying principle is similar to that of later approaches that were designed for social media data. The main steps of the proposed method include a) bursty features identification, b) bursty features grouping into bursty events, and c) determination of hot periods of the bursty events. The detection of bursty features is done using the probability of the number of the documents that contain a feature in a time window. This probability is modeled as a binomial distribution, of which the parameters are estimated from the whole corpus, i.e. all the time windows. For a specific time window, a feature is considered bursty if the number of documents  containing it is on the right side of the distribution. To group bursty features of a text stream into bursty events, a cost function that depends on the probability of bursty features grouped together is proposed. This cost function is used in a greedy fashion. 

The work of He et al. \cite{he2007analyzing} is similar to the previous approach, as it constructs a time-series signal for each term/feature. The signal is formed based on the DF-IDF values of each term per timeslot, where DF stands for the number of documents containing a specific word, while IDF for the inverse document frequency up to the current time window. The Discrete Fourier Transform is then applied on that signal and the dominant period and dominant power spectrum are determined. Based on these two values, each feature is classified as high or low power and high or low periodicity. Regarding periodicity, a signal with a dominant period higher than $T$/2 is termed 
aperiodic, otherwise it is defined as periodic. For the power spectrum, which expresses the overall activity of the feature in the corpus and indicates bursty features, the threshold is more difficult to set. In particular, a set of stop words is determined and the upper bound of these words’ dominant power spectrum is used as the threshold. Once each term is classified as periodic or aperiodic, the distribution of its appearance in time is modeled. For aperiodic features it is modeled using a single Gaussian, whereas for periodic events it is modeled using a mixture of Gaussian distributions. Then, in order to cluster the features, a cost function that depends on the KL-divergence between pairwise feature distributions and the number of documents in which both features appear is proposed. A greedy algorithm is used to minimize this cost function. 

TwitterMonitor \cite{mathioudakis2010twittermonitor}, one of the first feature-pivot approaches applied on social media messages, finds bursty keywords using an algorithm called  QueueBurst. This 
works with a single pass over the data. Once a set of bursty keywords are determined, they are grouped into trends by an algorithm called GroupBurst that is executed periodically. Not much detail is provided, however it is mentioned that it is based on the ``assessment of co-occurrences in recent tweets'' and that a greedy algorithm is used to perform the clustering in a computationally efficient manner. Subsequently, once such groups of terms are determined, TwitterMonitor uses context extraction algorithms like PCA and SVD over the recent history of trends to identify additional words that are associated with the trend but do not exhibit bursty behavior. Moreover, it detects the most frequently used external sources (URLs) and attempts to find geographical hotspots. It additionally uses Grapevine’s entity extractor to identify frequently mentioned entities in trends. 

The work of Sakaki et al. \cite{sakaki_www2010}, in which the authors mine tweets to detect events such as earthquakes and typhoons, is one of the pioneering works in event detection from Twitter. An SVM classifier, trained on a manually labeled Twitter dataset, is used to classify incoming tweets to the positive (earthquakes and typhoons events) or negative class (anything else). Three types of feature have been employed for this task: the number of words (statistical), the words in a tweet message, and the words before and after query words (contextual). Under the assumption that each Twitter user is regarded as a sensor, and each positive tweet as a sensory value targeting a specific event, the authors model event occurrences in time as an exponential distribution, while the location of earthquakes and trajectory of typhoon is estimated using spatial models based on Kalman and particle filters. Differently, to other feature-pivot methods, in this work the positive tweets constitute the only feature whose evolution is monitored for detection. 

Weng and Lee (2011) \cite{weng2011event} proposed the EDCoW method that uses the Discrete Wavelet Transform (DWT) on signals built from terms extracted from tweets. In contrast to the Fourier transform, which have been proposed in \cite{he2007analyzing}, DWT is localized both in time and frequency. The signal for each individual term extracted from tweets is built in two stages. In the first stage, the $DF \cdot IDF$ score of a word is calculated for each time point. $DF$ is the percentage of tweets containing the word in a time point compared to the number of all the tweets in the same period of time. $IDF$ is the inverse document frequency of a word from the first time-slot up the the current one. In the second stage, a sliding window is applied to capture the change in the frequency of a word over successive time windows. More specifically, the difference of the normalized wavelet entropy between the two time-slots is used to construct the second-stage signal. As a large number of words are trivial, auto-correlation is used to discard them on the basis of a dynamic threshold based on the median absolute deviation of all auto-correlations. The remaining words are then used to form a graph where similarity between them is calculated as the cross-correlation between the corresponding signals. Then, words are clustered to form events with a modularity-based graph partitioning technique, which splits the graph into sub-graphs each corresponding to a single event. Finally, to differentiate significant events from trivial ones, EDCoW also quantifies the events' significance using a score based on the number of words in the event and the sum of cross-correlation among them. 

Twevent \cite{li2012twevent} extracts consecutive non-overlapping segments from tweets, where a segment is defined either as a single word (unigram) or a phrase (multi-gram). Statistical information from the Microsoft Web N-gram Service and Wikipedia is used to detect non-trivial word segments by solving an optimization problem defined in the paper. Next, the frequency of each segment in a time-window is modeled as a binomial distribution, and bursty segments are defined to be those with frequency exceeding the expected value of each distribution. The bursty segments of a time window are ranked by their user frequency, which is the number of users who post tweets containing a segment during that time period and filtered further by keeping only the top-$k$ among them. A segment similarity that incorporates the content of the associated tweets and the temporal pattern of segments within the specified time window is used to construct a graph of segments. Then a variant of the Jarvis-Patrick graph clustering algorithm is applied to group related segments. Finally, the detected events are ranked and filtered based on a newsworthiness score calculated using Wikipedia. 

EnBlogue \cite{alvanaki2012see} detects topics emerging in streams of news documents, blog posts and tweets by detecting abnormal shifts in correlations of tag pairs within a given window of time.  In other words, the feature used for the detection of events takes the form of pairs of tags. Initially, seed tags are identified by detecting popular tags compared to a sliding window average, and only these are consider in the next steps. These seed tags are used to generate candidate topics, i.e. pairs of tags that contain at least one seed tag.  For each such pair the correlation between the two tags is calculated over time, based on the amount of documents that are annotated with both tags in each timeslot. For each candidate pair, the previous correlation values are used to predict the correlation in the current timeslot. If the actual correlation is larger than the predicted value then the topic is considered to be an emerging topic and the relative prediction error is used as a ranking criterion. For prediction, exponential smoothing is used, which is a forecasting technique that uses a weighted moving average of past data as the basis for the forecast.

The event detection method proposed by Cataldi et al. \cite{cataldi2010emerging} models the lifecycle of terms according to a novel aging theory based on user authority and uses that lifecycle to select  emerging terms. Initially, the method creates a representative term vector for each tweet within a given time window. The term weights are calculated using the augmented normalized term frequency weighting scheme. In the next step, the authority of users is calculated by using a follower graph, following a PageRank-like approach. The proposed content aging theory is then used to detect the lifecycle of each of the keywords. Given a keyword appearing in a specific time interval, a formula, termed as nutrition formula, evaluates the usage of this keyword by considering its frequency in the tweets that mention it and also the authority of each single user that reports it. Once the nutrition of a keyword is obtained for the current timeslot, its energy in the same timeslot is calculated based on the nutrition values of the current and the previous timeslots. Next, the authors propose an approach for the selection of emerging topics based on thresholding of their content energy, where a critical drop value is calculated dynamically based on the average energy value of all keywords. Every keyword with energy below that drop value is discarded, while the remaining keywords form the set of emerging keywords in that timeslot.  Each emerging keyword is associated with a correlation vector formed by a set of weighted terms that defines the relationships among that keyword and all other emerging keywords in the considered time interval. Correlation vectors are used then to generate a graph of keywords,  and topics are identified by detecting  the strongly connected components of the graph. 

Parikh and Karlapalem \cite{parikh2013events} proposed an event detection system targeting Twitter, termed ET. 
The first step of ET aims at extracting event representative features: those are features exhibiting a significant increase in their appearance in a particular timeslot compared to the previous ones. Features correspond to bi-grams after the removal of stop-words and a list of frequent timeslots is associated with each of the features. Next, the set of event representative keywords, is clustered in groups related to the same event by using a combination of content similarity and appearance pattern similarity. The content similarity between two features is calculated by finding the Jaccard similarity between the sets of their associated tweets, but only for the timeslots  that the features are frequent. The appearance pattern similarity is the Jaccard  similarity between the  frequent timeslots of the features. The overall similarity is the weighted sum of the above two similarities, and an agglomerative hierarchical clustering technique is used to group features into events.  

The approach of Chen \& Roy \cite{chen2009event} is one of the first to detect events in multimedia collections by using feature-pivot methods in the associated tags. Given a set of Flickr photos associated with user provided tags and other metadata, including time and location, the objective is to discover a set of photo groups, where each group corresponds to an event. Associated with photos, each tag occurrence can be modeled with temporal and spatial distributions. These distributions of tags occurrences are analyzed and events are identified as groups of tags that exhibit a similar temporal and spatial behavior. In this work DWT is employed to discover event-related tags with significant distribution in both dimensions. Due to data sparsity, the authors propose the quantization of the data in the original 3D space before the application of DWT. More specifically, the 3D space is segmented into cells by dividing each dimension into intervals of equal size. For the latitude and longitude dimensions, the interval size is set to 1, while for the time dimension each interval represents one day. Next, DWT is applied on each dimension and the cells with weak wavelet coefficients in the transformed space are removed. Dense regions in the transformed wavelet space are detected and propagated from the transformed space to the original spatio-temporal space. Tags that do not belong to any significant region are removed as they are unlikely to be related to events. Finally,  to cluster event-related tags a combination of semantic and spatial similarity is used in conjunction with DBSCAN as the clustering method. 

Sayyadi et al. \cite{sayyadi2009event} present a feature-pivot approach termed KeyGraph. For each term extracted from documents, its DF, IDF and term frequency are calculated for each timeslot. Then, a KeyGraph is generated by creating a node for each term with DF above some threshold. Terms that co-appear in a number of documents above a threshold are connected. If the conditional probability of the one appearing given that the other appears in some document is below some threshold, the corresponding edge is filtered. The nodes of the resulting KeyGraph are clustered using a community detection algorithm. The algorithm uses the betweenness centrality of edges. Betweenness centrality quantifies the number of shortest paths between any pair of nodes in the graph that pass through a particular node/edge. Nodes/edges with high betweenness centrality values are likely to connect different communities, parts of the graph that are more densely connected inside them than to the rest of the graph. The algorithm progressively removes  edges with higher betweenness centrality and that results in the individual communities of the graph to separate from each other. Additionally, since a term may be relevant to more than one event, the adjacent nodes of a pruned edge, are duplicated in the two sides of the pruned edge (if the conditional probability of the nodes is above some threshold). Finally, for each cluster, a key document is formed by taking the set of terms that have been assigned to it. The set of actual documents are assigned to the clusters by computing cosine similarities. Regarding implementation, in order to more effectively compute betweenness centrality, the authors propose to sample pairs of nodes in order to compute shortest paths. Also, to deal with streaming data, the authors propose the periodic execution of the algorithm on overlapping windows. Topics that are common on subsequent executions can be determined and merged by looking at the documents that are common on both windows. 

A different feature-pivot approach is that of Guille and Favre \cite{guille2014mention}, in which the role of features is played by terms contained in tweets with mentions between Twitter users. As stated by the authors, in most existing feature-pivot methods, the focus is on features extracted from the textual content of tweets, while the social aspect of social media platforms is ignored. 
The system proposed in this work consists of successive phases for the detection of events, the selection of representative words describing the events, and finally their ranking based on impact and significance. Regarding the event detection phase, the method considers terms contained in tweets with user mentions, and detects whether a term exhibits a burst at each time point by modeling its appearance as a binomial distribution. To identify the interval of the burst for each term, a maximization problem is solved permitting the identification of events of arbitrary length. The identified burst terms are considered as candidate events, and are annotated with additional terms that describe the event by selecting terms having similar temporal dynamics with the main term. 

Zhang et al. \cite{zhang2015event} proposed a system to detect burst novel events and predict their future popularity. For event detection term bursts are detected, where a term burst in a given time window is identified by examining the increase in the term weight in that window. The weight of a term in a micro-blog post is calculated as the combination of the augmented normalized term frequency and authority of the user that posted it. User authority, estimated by the PageRank algorithm, is incorporated into term weighting under the assumption that the use of a term from influential users increases the significance of the term. To detect bursts, the term usage in each time window is modeled as a Hidden Markov state model (HMM) with two states, ``low'' and ``high''. From the detected bursty terms in a time window, a directed graph is constructed, where vertices correspond to the terms, and edges between them represent term co-occurrences. The weight of an edge between two terms is estimated using the set of micro-blogs containing both terms as positive evidence, and the set of micro-blogs containing only one of them as a negative evidence against their relation. As that relation is asymmetric, two directed edges are added for each pair, while a normalization method is used to remove edges with small weights. Next, the strongly connected components are detected to form event clusters. 

\subsection{Document-pivot methods}
\label{sec:doc_pivot}

This section discusses methods that detect events by clustering documents on the basis of their semantic similarity. Document-pivot approaches, originating from the field of Topic Detection and Tracking task (TDT) \cite{allan2002introduction} can be seen as a clustering problem. Both RED and FSD approaches have been proposed. In both cases, the underlying principle is quite similar: documents are modeled in ways that capture their semantic content and then a clustering algorithm is applied to group them into events. What differs between approaches is mainly the characteristics of the clustering approach, the way that documents are mapped to feature vectors, and the similarity metric used to identify whether two documents are from the same event. For example, in retrospective event detection, the whole dataset is available, thus an a priori knowledge of the documents' characteristics is available. Compared to the traditional TDT task that assumes that all documents are relevant to events, in social media data, event-related data are mixed with noisy data. In addition, the amount of social media data is much larger than the typical size of a dataset in traditional TDT. Thus, there is a need for scalable and efficient methods. Also, in contrast to TDT where items are text documents, social media items are multi-modal in nature, and often contain visual content, as well as temporal and spatial information. So, their representation is different from the traditional Bag-of-Words representation of documents used in most TDT approaches. 

\begin{figure}
\centering
\includegraphics[width=1.02\textwidth, trim={0 2cm 0 1.5cm},clip]{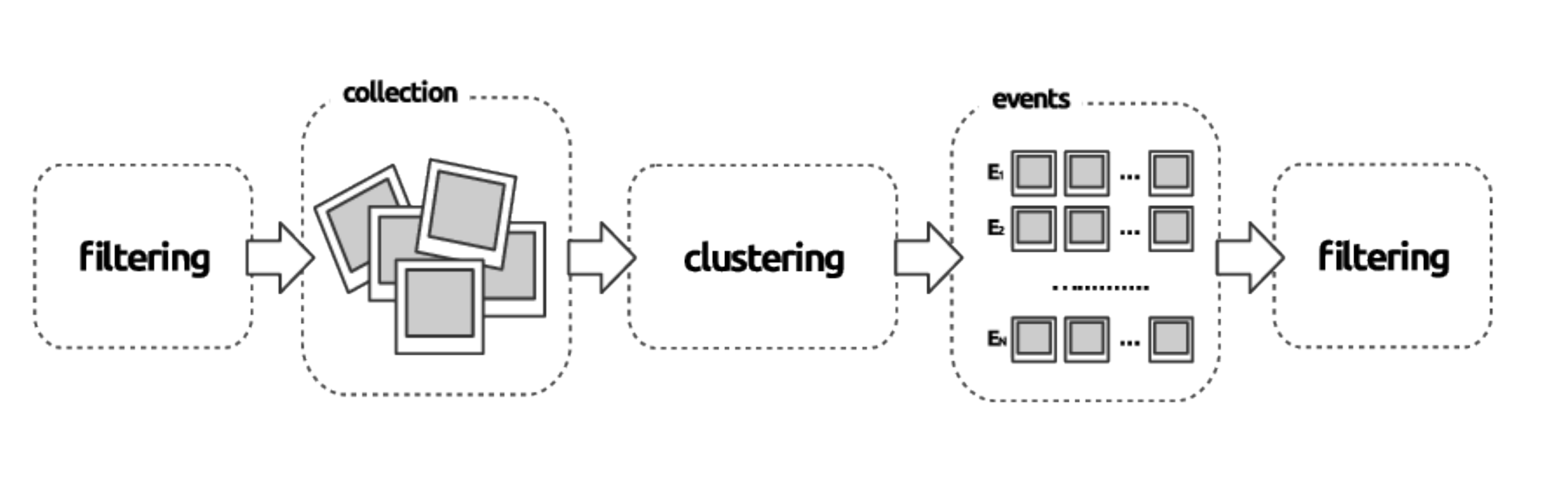}
\caption{Document-pivot event detection}
\label{fig:doc_pivot_detection}
\end{figure}

TwitterStand \cite{sankaranarayanan2009twitterstand} was proposed as a news processing system based on Twitter for capturing tweets corresponding to breaking news and for organizing them around these news. The input to TwitterStand ranges from a set of seeders, which are handpicked users that are known to publish news. 
As collected tweets are noisy, the authors apply a Naive Bayes classifier to classify incoming tweets as junk or news, where junk tweets have a high probability of not being related to the news and hence, are discarded. Tweets are represented by a feature vector using the TF-IDF weighting scheme and clustered into news topics by applying an online clustering algorithm, called leader-follower clustering, which allows for clustering in both content and time. Unsurprisingly, the proposed method has low performance due to fragmentation of news topics, noisy content and the limited power of Bag-of-Words in short texts as tweets. To this end, the authors suggest additional improvements to ensure better clustering results. 

The work of Petrovic et al. \cite{petrovic2010streaming} focuses on the FSD problem, which is closely related to the new event detection task. 
This is done by finding the closest neighbor of an incoming document. If the similarity is below some predefined threshold, meaning that no such neighbor exists, a new story is detected and added to the collection of stories. The main contribution of this work is that the authors propose the use of Locality Sensitive Hashing (LSH) to retrieve fast the nearest neighbor of the incoming tweets. Tweets are represented as term features vectors and indexed using LSH. However, LSH, as an approximate method, finds the nearest neighbor only if it is reasonably close to the query. The authors propose to compare the distance of the nearest neighbor with some threshold and if it is above some threshold to execute a linear search on the $N$ most recently arrived documents. Additionally, in order to overcome the problems that appear when dealing with open ended data streams, only the most recent documents are considered. Moreover, the authors discuss the problem of event detection: if the best matching document/cluster has similarity above some threshold, then the new item is added to the cluster, otherwise a new cluster is created. Finally, different ways of ranking and filtering the detected clusters are introduced based on the number of tweets or users as well as on the entropy of the number of users.

Becker et al. \cite{becker2011beyond} focus on clustering event-related tweets on the same cluster and mainly on separating clusters of tweets as either representing a real-world event or not. In order to group similar tweets, a simple incremental online clustering algorithm is used. Each tweet is represented as a TF-IDF vector and each cluster is represented as the average TF-IDF vector of the tweets that it contains. When a new tweet arrives, its cosine similarity to all existing clusters is computed. If the similarity to the best matching cluster is above some threshold, then the tweet is inserted into the cluster, otherwise, a new cluster is created. The main focus of that work is the classification of clusters in Event and Non-Event classes. Compared to other works such as \cite{sankaranarayanan2009twitterstand}, this work attempts to discard irrelevant information at the level of clusters instead of doing that at the level of individual items. For each cluster the following features are computed: 1) Temporal features. Two temporal features are used. The first is the deviation from the expected number of tweets containing some term given the frequency of appearance of the term in the last $N$ time units. The second examines how well the number of term appearances fits an exponential growth pattern. These features are computed for the $n$ most frequent terms of each cluster. 2) Social features. The number of social interactions that appear in the tweets (number of mentions, re-tweets and replies). 3) Topical features. Three different features are used, namely a) the average or median similarity of messages to the cluster centroid, b) the percentage of tweets containing the first, second, etc. most frequent term, and c) how many of the most frequent terms are contained in at least $n\%$ of the the tweets. 4) Twitter centric features. Three features are used, namely a) the percentage of tweets that contain tags, b) the percentage of tweets that contain the most frequently used tag, c) a binary variable indicating whether the most frequently used tag is a concatenation of multiple words or not.

Lee \cite{lee2012mining} proposed to apply a density-based online clustering method on micro-blog text streams to obtain temporal and spatial features of real-world events. Regarding the representation of incoming documents, a novel term weighting method, called \textit{BursT}, is used. This method is supposed to handle the concept drift problem that occurs in evolving data streams. The similarity between two documents is defined as the cosine similarity of the corresponding BursT vectors, and a temporal penalty is introduced to adjust the similarity according to the time distance between documents. To deal with the large volume of documents, the authors propose the use of a sliding window model, i.e. only a subset of the most recent documents is kept for analysis. Finally, the employed clustering algorithm is the density-based incremental DBSCAN. Therefore, the shape of clusters may change over time. When event clusters are detected, the next step is to analyze their spatial distribution. A location feature vector, which records the location distribution of the events at specific time points, is used to distinguish whether the detected events are local or global. 

A major challenge that the previous approaches face is the high degree of lexical variation in documents that makes it difficult to detect stories about the same event using different words. The problem becomes even worse in micro-blog platforms such as Twitter given the brevity (limited text content) of posts. In that scenario, a tweet could be flagged as a first story even if a related tweet, which uses different but synonymous words, was already returned as a first story. Petrovi\'{c} et al. 2012 \cite{petrovic2012using} extended their LSH-based method \cite{petrovic2010streaming} by suggesting the use of paraphrases to alleviate this problem. They proposed a novel way of integrating paraphrases with LSH in order to obtain an efficient FSD system that can scale to very large datasets. In a similar way, Moran et al. \cite{moran2016enhancing} proposed a new way of automatically computing the lexical paraphrases using word embeddings. There has been an extensive amount of prior research that has shown that the cosine similarity between word embeddings is correlated with the semantic relatedness between the corresponding words. Moran et al. made use of this property by considering two words to be lexical paraphrases if the cosine similarity between their word embeddings is sufficiently high. After that step, the same approach the one by Petrovi\'{c} et al.  \cite{petrovic2012using} is followed.  

Aggarwal and Subbian \cite{aggarwal2012event} propose an incremental online clustering approach to detect events from social streams. Clusters are represented with cluster summaries, which are used for comparisons with the incoming documents. Each  summary, corresponding to an event, contains a node-summary, which is the set of users together with their frequencies and a content-summary, which is a set of words and their corresponding TF-IDF values. To maintain and update these summaries per cluster, a sketch-based technique is used. The similarity between a new document and a cluster is the combination of nodes and content similarity, controlled by a balancing parameter. By incorporating nodes similarity, the authors propose a novel similarity that exploits the underlying social structure of the network. According to them, this improves the content-based similarity proposed by previous works \cite{becker2011beyond,petrovic2010streaming,sankaranarayanan2009twitterstand}. Each incoming document is assigned to its closest cluster, according to the aforementioned similarity, unless this similarity is significantly lower than that of other documents assigned to the cluster. A similarity score is considered significantly lower if it is lower than $\mu-3\sigma$, where $\mu$ is the mean value of all previous similarities, and $\sigma$ is the standard deviation. In that case, the document creates a cluster of its own, and one of the current clusters is removed on grounds of staleness. In that way, by keeping a fixed number of cluster summaries, the comparisons required is reduced to a fixed number.  

Becker et al. \cite{becker2009event} address the problem of partitioning a set of documents into clusters in a way that each cluster corresponds to the documents that are associated with one event. They use multiple types of social media features, including text, time and location. For each modality, an appropriate similarity function is used. For example, cosine similarity for text and Haversine distance for location. The paper proposes ensemble clustering in conjunction with online clustering algorithm, as a method to cluster similar documents into events.  Ensemble clustering is an approach that combines multiple clustering solutions for a documents set. The first step of ensemble clustering is to select a set of \textit{clusterers} that partition the data using the different modalities and the appropriate similarity metrics. To cluster the document according to each modality, the authors used a single-pass incremental algorithm, similarly to \cite{becker2009event}. More precisely, given a threshold $\mu$, a similarity function according to some modality and a set of documents, the algorithm assigns each document to the best matching cluster if the similarity is above $\mu$. Otherwise a new cluster is created. A training step is necessary to select the most suitable threshold setting for each modality and to assign a weight to each clusterer indicating the confidence in its predictions. The second step  combines individual partitions to a single partition. Each clusterer is regarded as providing an expert vote on whether two documents belong to the same cluster. For a pair of documents and a clusterer, a prediction function is defined as equal to 1 if the two documents are in the same cluster, and 0 otherwise. Then, the consensus score for that pair of documents is computed as the weighted sum across all clusterers. This is a late fusion strategy, where the different modalities are combined at the last stage of the procedure. 

There are many research efforts in the direction of early fusion, by integrating different modalities in a single similarity metric. The work of Becker et al. \cite{becker2010learning} extends their previous approach \cite{becker2009event} by proposing alternative ways of combining the results of ensemble clustering. Instead of combining individual partitions as in traditional ensemble clustering setting, the learned weights and thresholds are used as a model for similarity metric across multimodal social media items or between items and cluster centroids. Hence, a single-pass incremental clustering algorithm can be used in conjunction with the combined similarity metric. The authors also propose the use of a classification model to learn document similarity functions for social media. In other words, a classifier using as input features the raw similarity scores corresponding to the document features is used to compute whether a pair of documents belongs to the same event. Both SVM and Logistic Regression have been used, with the latter experimentally shown to perform better. 

\begin{figure}
\centering
\includegraphics[width=0.9\textwidth, clip]{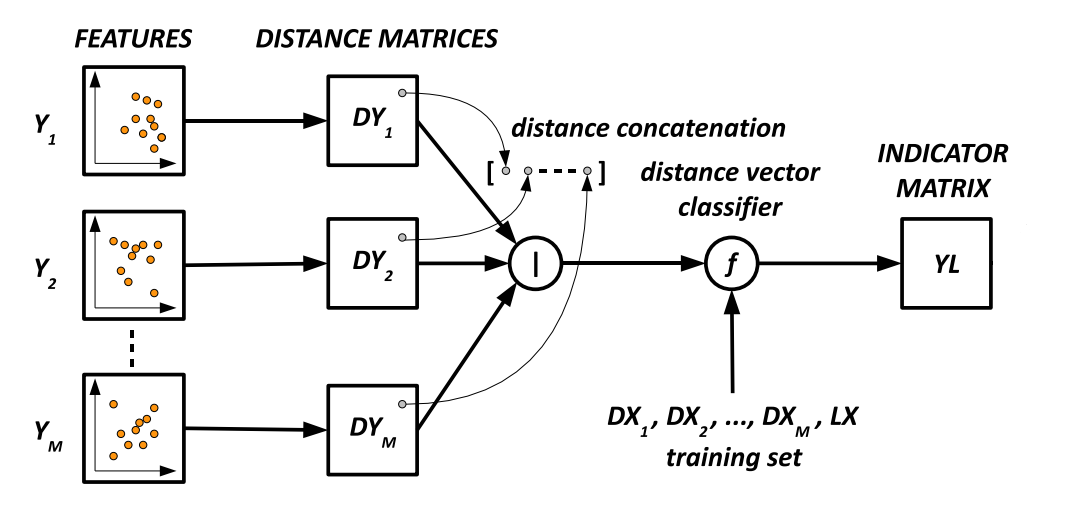}
\caption{Similarity learning approach by Petkos et al. \cite{petkos2012social}. An annotated dataset $X$ is used to obtain training examples consisting of distance vectors and the corresponding labels. These are used to learn the classifier $f$ that is applied on the distance vectors of a new dataset $Y$ to predict the corresponding indicator vectors that are used for clustering. 
}
\label{fig:sim_learning}
\end{figure}

Reuter \& Cimiano \cite{reuter2012event} attempt to solve the problem of grouping documents related with the same event into the same cluster in a similar manner with \cite{becker2010learning}. The authors  use a single-pass incremental clustering algorithm and define the similarity between an incoming document and a given event cluster as the likelihood probability that the document belongs to the cluster. This probability is computed using an SVM classifier on the textual, geographical and temporal dissimilarities between documents and clusters. Two contributions of \cite{reuter2012event} compared to \cite{becker2010learning} is the introduction of a candidate retrieval step and the use of an additional classifier that determines whether the incoming document belongs to the top scoring candidate event or rather to a new event. The candidate retrieval step aims to make the method scalable by decreasing the number of comparison between incoming documents and events. Specifically, for each document the nearest events are retrieved with respect to each modality. For the new event detection problem, instead of using a simple threshold and assign a document to an event if the similarity exceeds this threshold, the authors propose the use of a classifier that predicts the probability that a document belongs to an existing or to a new event. 

Petkos et al. \cite{petkos2012social} present an approach for multimodal clustering designed for Flickr images. First, a baseline approach is presented, in which a weighted sum of the affinity matrices of the different modalities is computed and then, similarly to a common spectral clustering algorithm, the corresponding graph is pruned, the Laplacian matrix is computed and the minimum cut of the graph is found. Given the difficulty to estimate the best values for the different parameters of that approach, and the fact that the clustering output is sensitive to these parameters, an approach that uses a known clustering is proposed to supervise the multimodal fusion and clustering procedure. Pairwise distances per modality are first computed. Then, for each pair of items, the distances per modality are compiled in a single vector. These dissimilarities  vectors is used to learn the typical ``same cluster'' relationship for multimodal items. Having learned such a classifier, one can form an indicator vector for each of the new items to be clustered. This indicator vector summarizes the ``same cluster'' relationship between each item and all other items to be clustered. Items that have similar indicator vectors should belong to the same cluster. Therefore, a final clustering result can be obtained by clustering the indicator vectors, for example by applying a spectral clustering algorithm.

Bao et al. \cite{bao2013social} proposed a method for the clustering of social multimedia into events without the need for a single representation for each individual item. Contrary to other methods  that adopt early or late fusion for combining the different modalities, the authors represent a collection of social media items as a star-structured $K$-partite graph, where the items/documents are regarded as the central vertices and different modalities are treated as the auxiliary vertices that are pairwise independent with each other but correlated with the central vertices. Four types of modalities are considered, publication time, location, visual content and text, while different ways to calculate inter-relationships are used. Then, a graph co-clustering approach, termed information theoretic co-clustering, is used to partition the graph into events. 

Wang et al. \cite{wang2012social} proposed an event detection method 
that incorporates online social interaction features. 
As stated by the authors, as real-world events are intrinsically social, it is natural to expect that participation in the physical event will motivate people attending them to interact online with shared media associated with that event. Social interactions around photos, including actions such as posting a comment, adding a tag, or liking the photo, are modeled using an undirected graph with three conceptual layers: photos, users and tags. Each unique user, photo and tag form the vertices of the graph, while edges, symmetric or asymmetric, can have various semantics. The social affinity between two photos is the random walk probability from one photo to the other, calculated by a Random Walk with Restart algorithm. This affinity, combined with other similarities from features like time, location and text, is used in an SVM classifier to calculate a single value similarity measure, as in \cite{becker2010learning} and \cite{reuter2012event}.

Petkos et al. \cite{petkos_mtap2017} recently proposed a clustering approach for the multimodal clustering problem. Similar to their previous work \cite{petkos2012social}, relevant clustering examples are used to learn an SVM classifier of the ``same cluster'' relationship between pairs of multimodal items. 
The used modalities include text, visual content, the uploader's identity, creation time and location. An important component of the proposed approach relies on the capability for fast retrieval of same cluster candidates per item for each of the different modalities, so that the model can be applied only on the candidate items. For that purpose a different index was used per modality. Given two images, which are expressed through a set of features, a vector that contains the per-modality distances is calculated. The same cluster model then predicts if the two items belong to the same cluster. 
The learned model is subsequently used to organize items in a graph, where the nodes represent the items and links between pairs of nodes are inserted if the model predicts that the corresponding items belong to the same cluster. Eventually, a graph clustering algorithm is applied to produce the final clustering. 
The proposed approach uses two variants, a batch community detection algorithm (SCAN) and an incremental algorithm, the Quick Community Adaptation (QCA).

\subsection{Topic modeling methods}
\label{sec:topic_models}

\begin{figure}
\subfloat[Latent Dirichlet Allocation (LDA) \cite{blei2002latent} ]{\includegraphics[width= 1.0in]{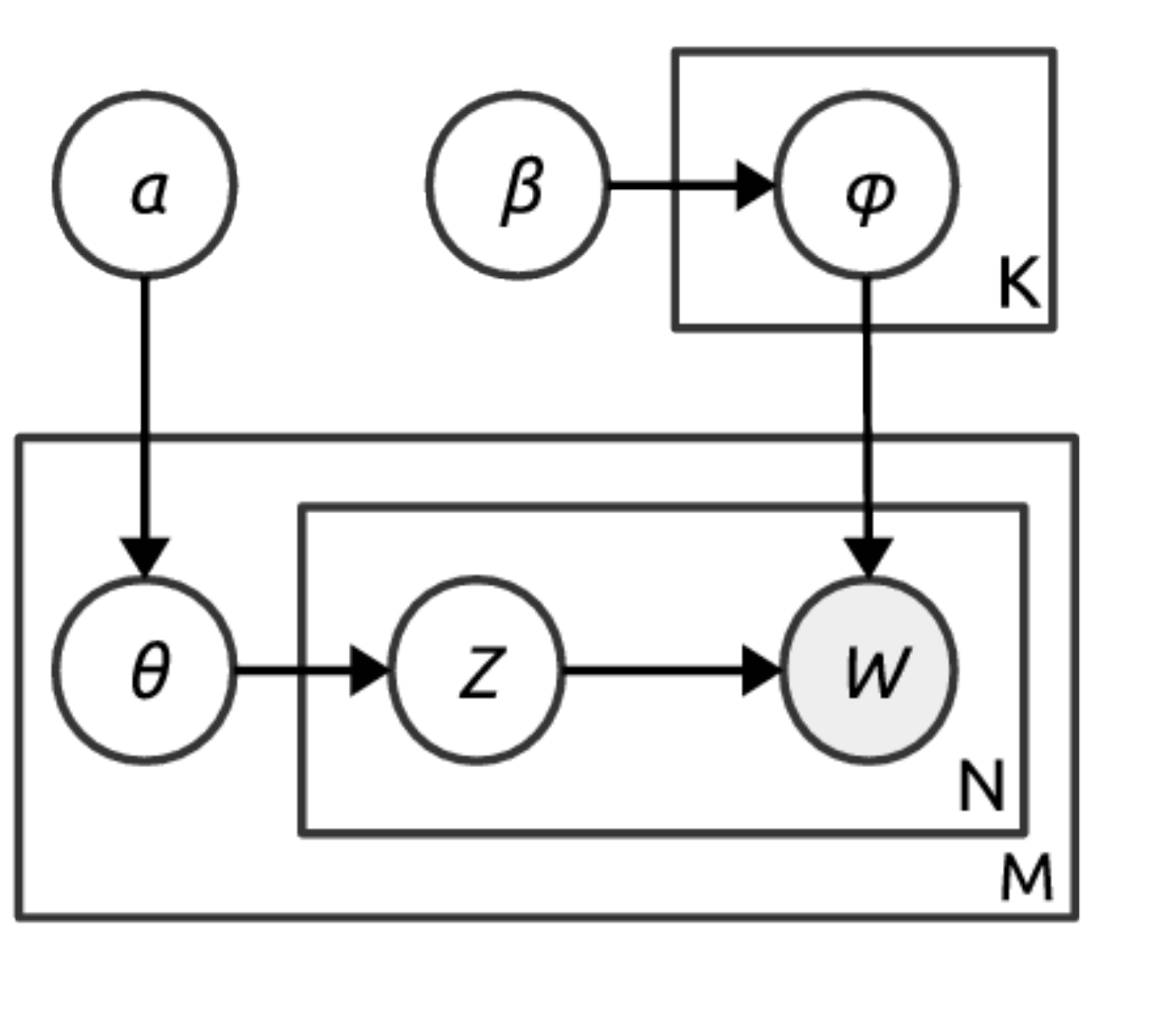}\label{fig:lda}}\hspace{0.2cm}
\subfloat[Twitter LDA \cite{zhao2011comparing}]{\includegraphics[width= 1.2in]{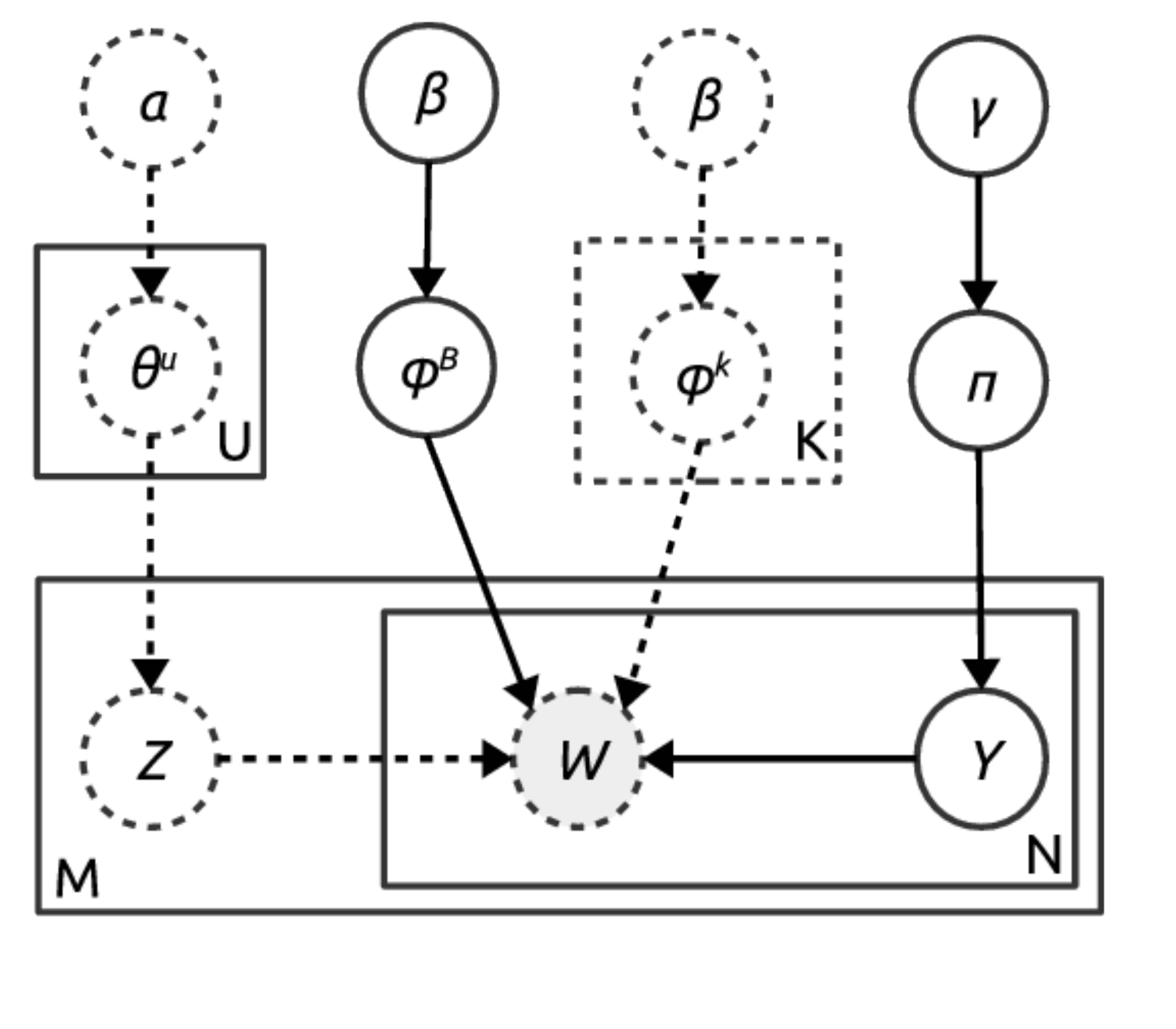}\label{fig:twitterlda}}\hspace{0.2cm}
\subfloat[Topic over Time model (TOT) \cite{wang2006topics}]{\includegraphics[width= 1.1in]{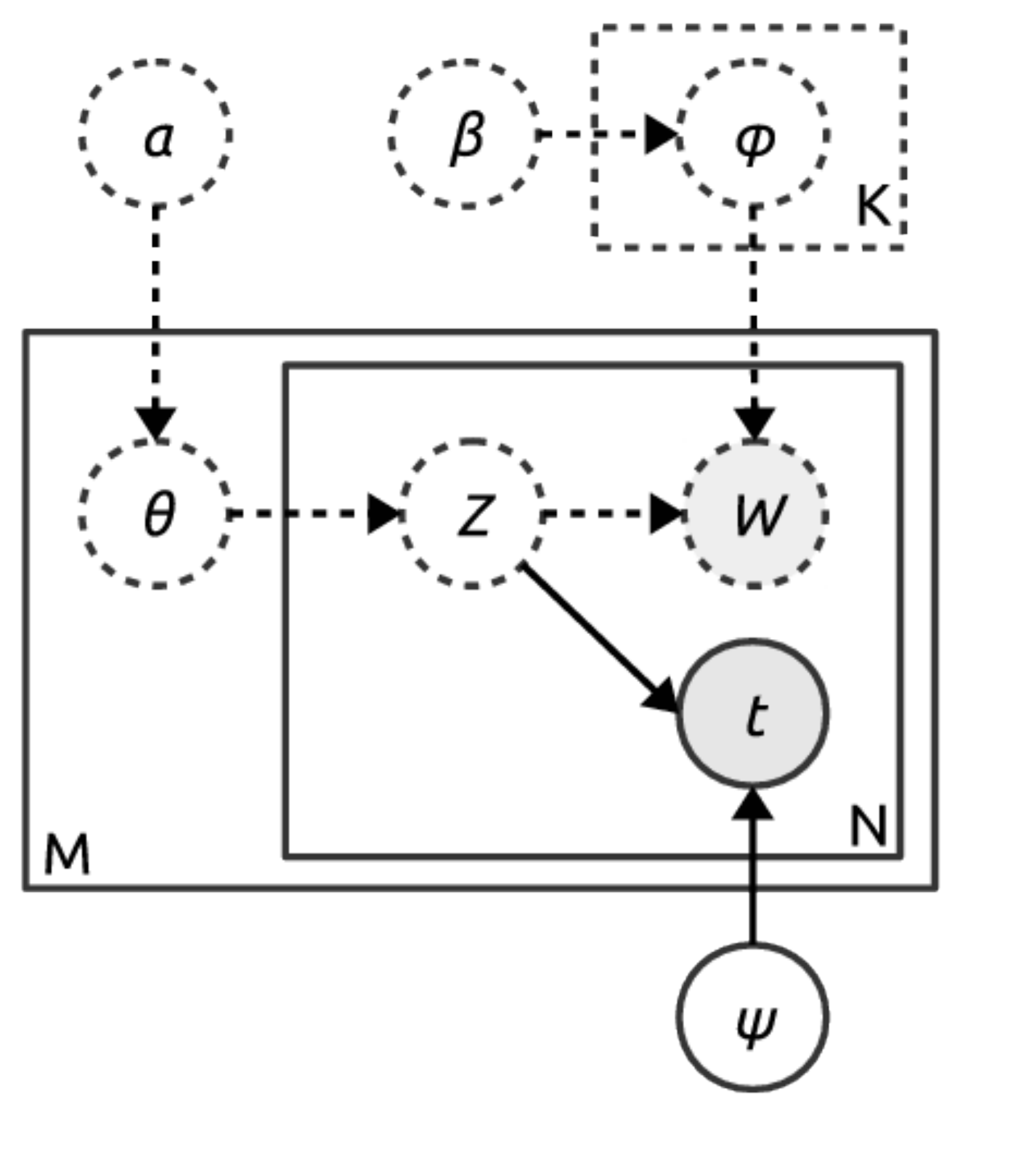}\label{fig:tot}}\hspace{0.2cm}
\subfloat[Location Time Constrained Topic (LTT) \cite{zhou2014event}]{\includegraphics[width= 1.in]{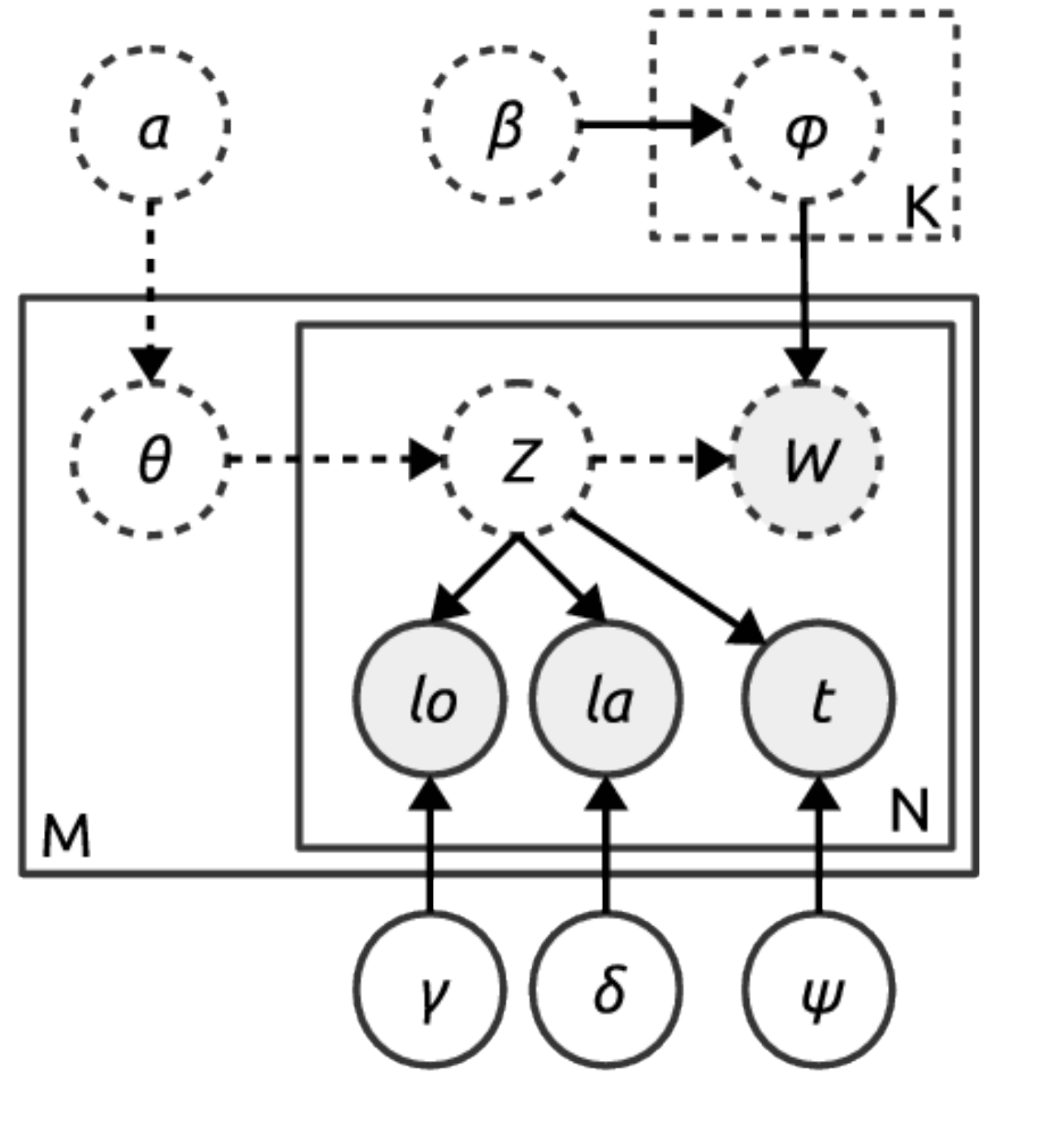}\label{fig:ltt}}\\
\subfloat[MultiModal LDA (MMLDA) \cite{bian2013multimedia}]{\includegraphics[width= 1.4in]{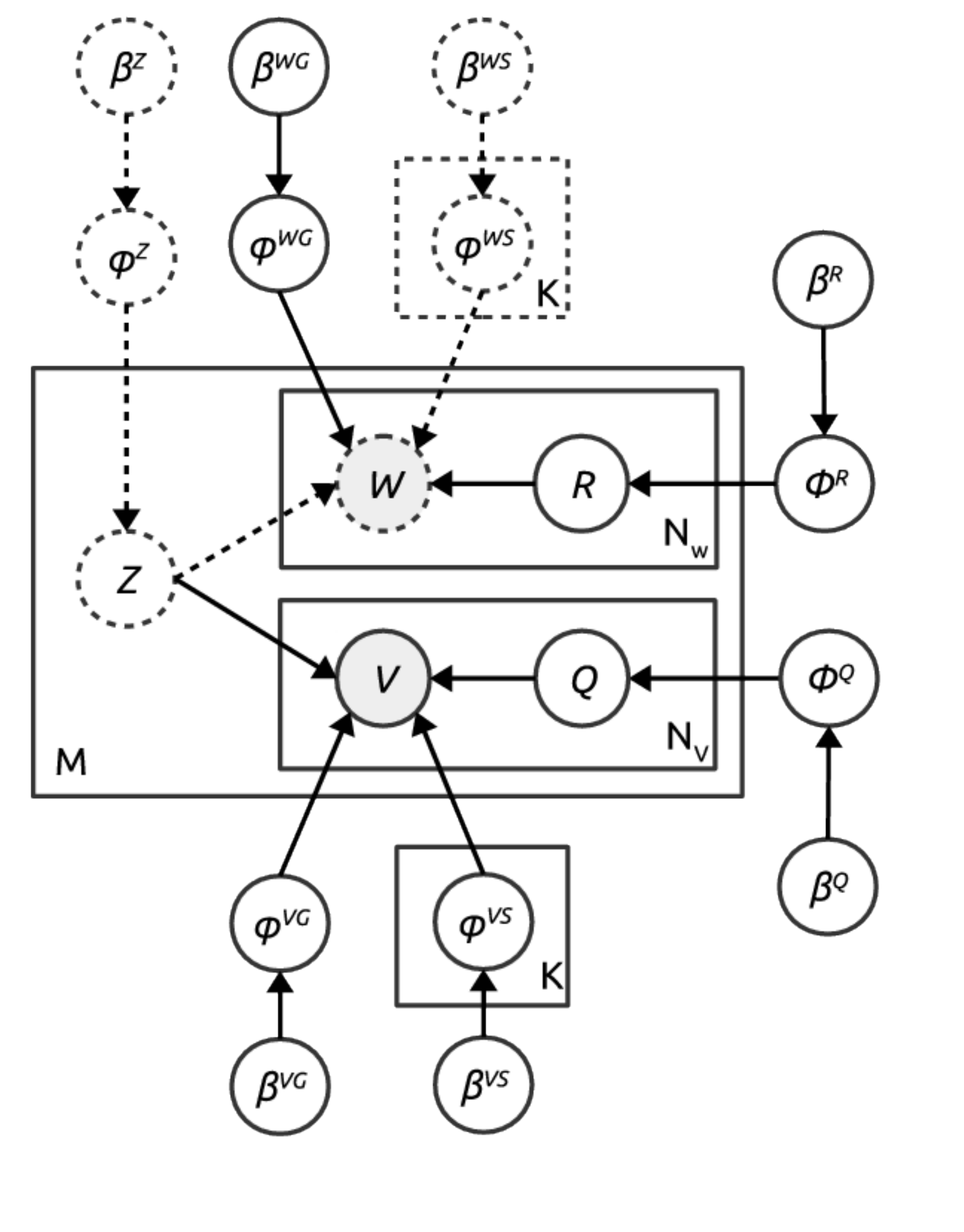}\label{fig:mmlda}}\hspace{0.3cm} 
\subfloat[Bayesian Graphical Model for Latent Events discovery \cite{wei2015bayesian}]{\includegraphics[width= 1.5in]{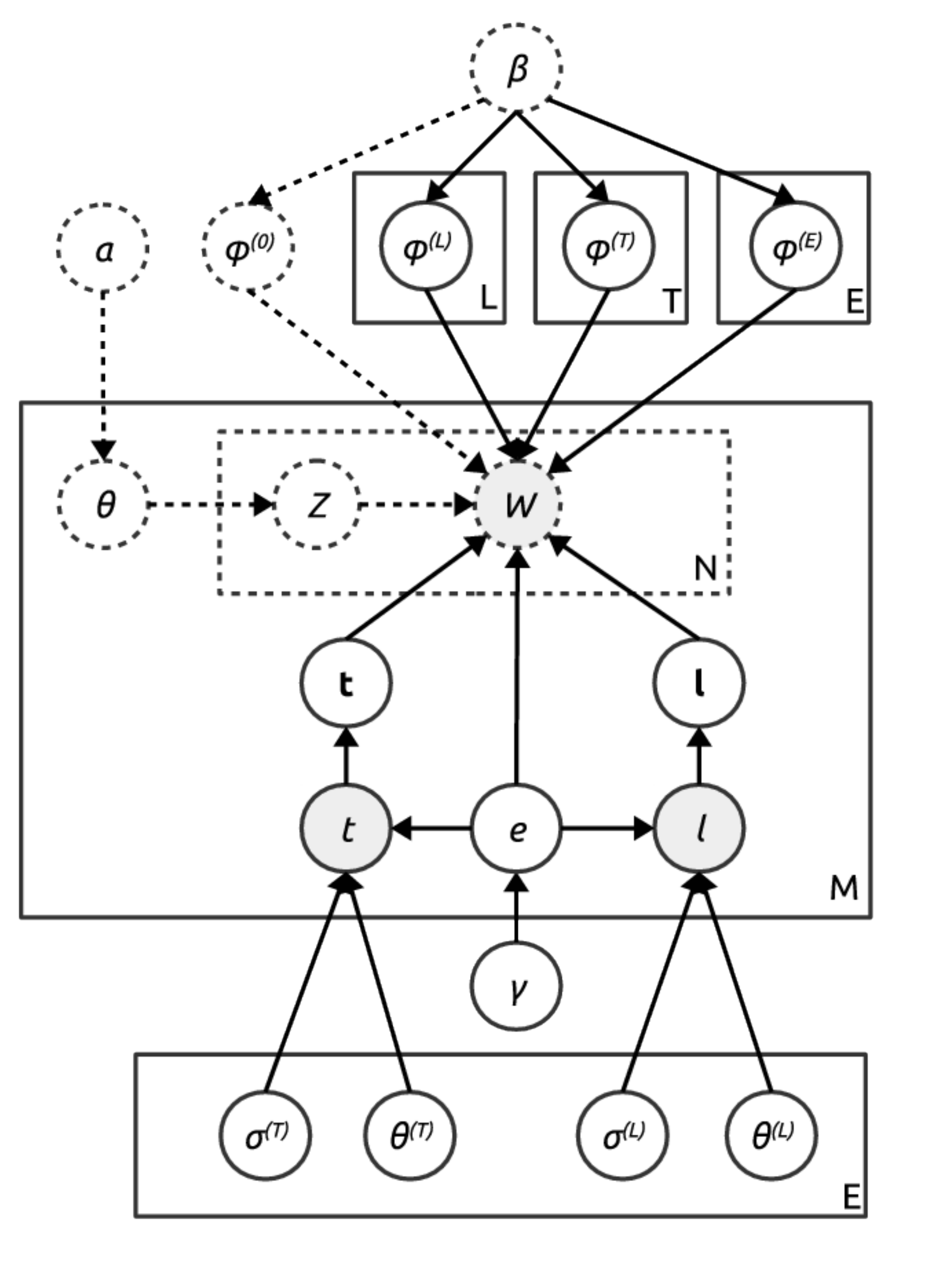}\label{fig:bgm}}\hspace{0.3cm}
\subfloat[Spatio Temporal MultiModal TwitterLDA (STM-TwitterLDA) \cite{cai2015popular}]{\includegraphics[width= 1.6in]{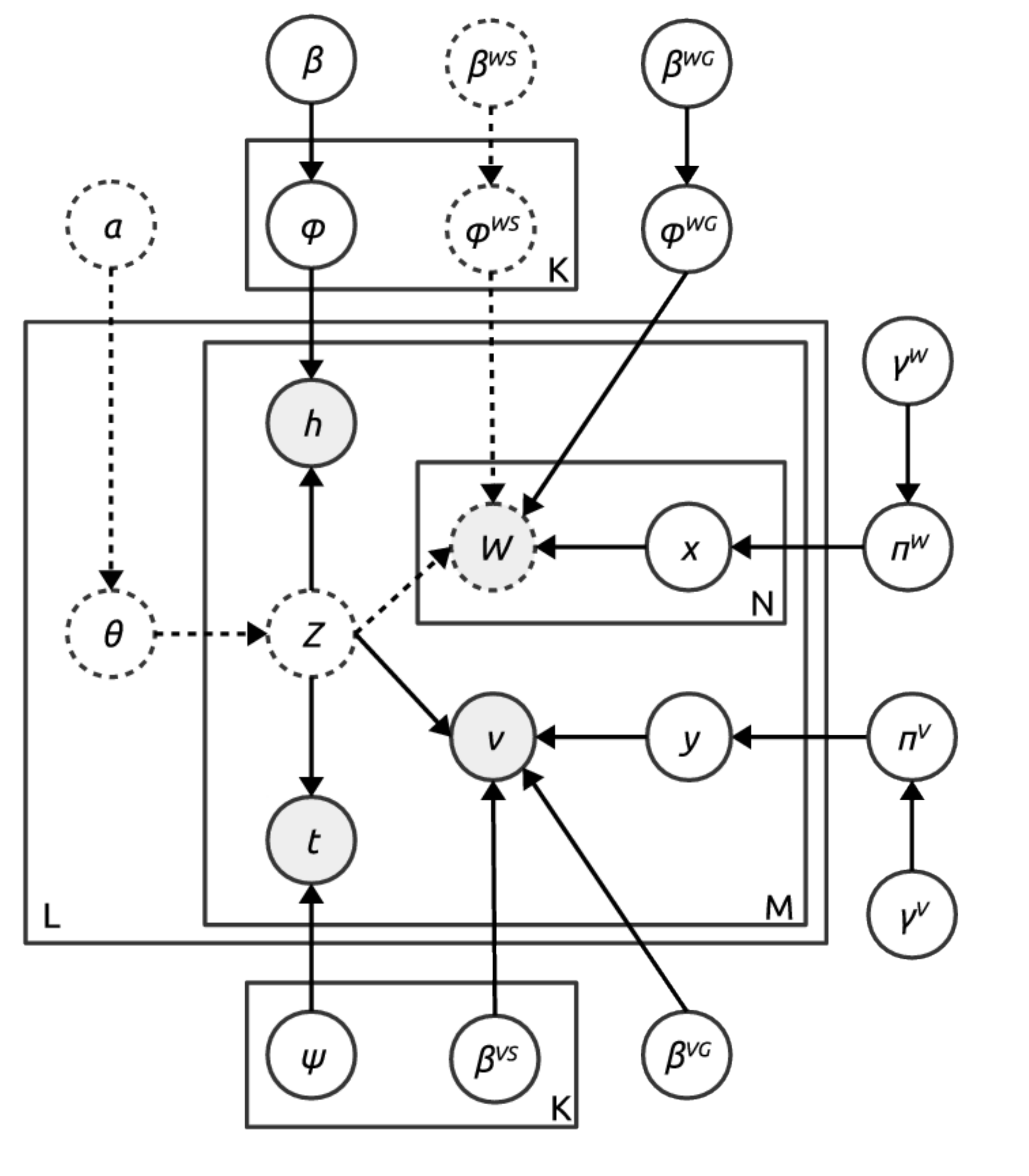}\label{fig:stm}}\\
\subfloat[General and Event related Aspects Model (GEAM) \cite{you2013geam}]{\includegraphics[width= 1.1in]{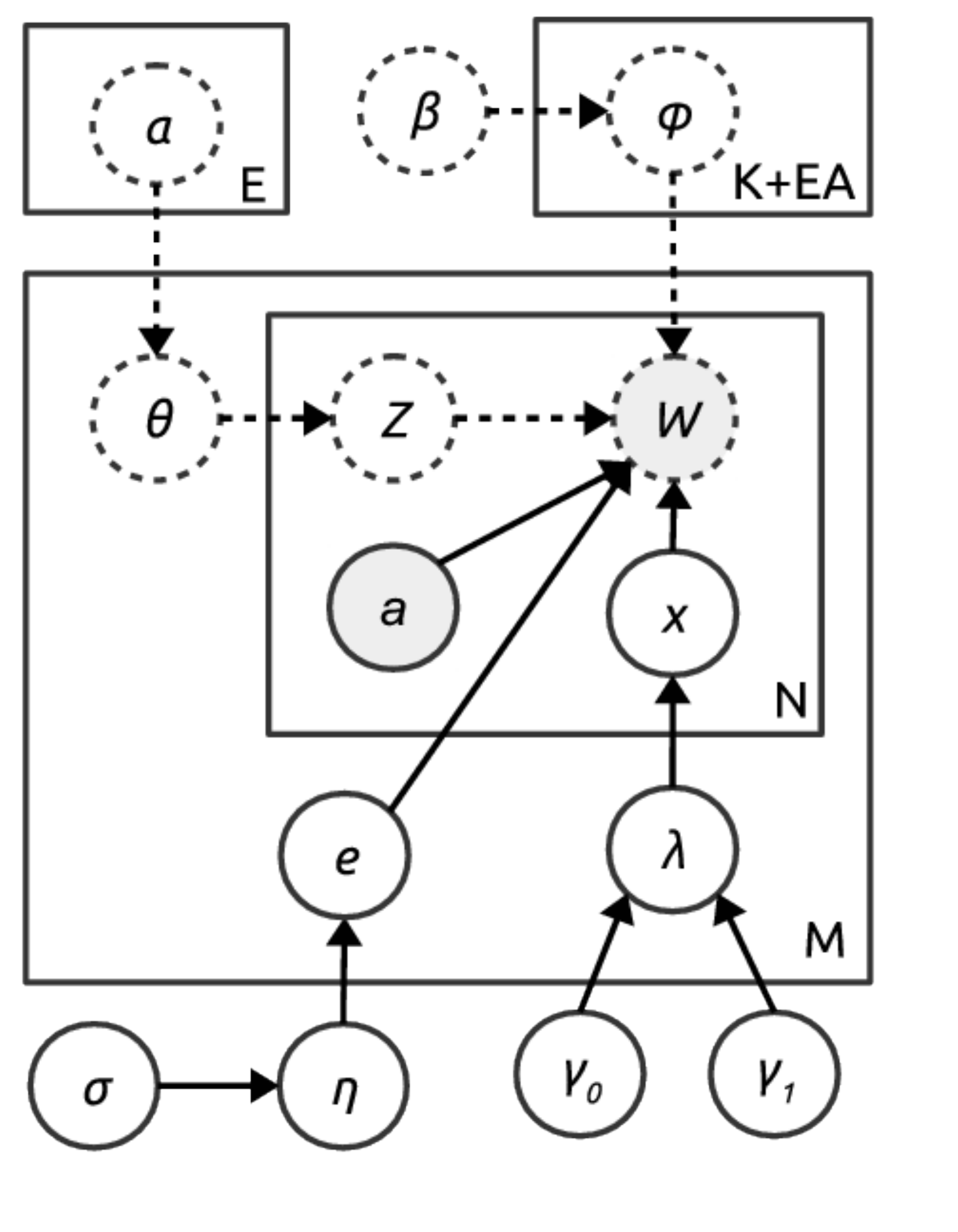}\label{fig:geam}} \hspace{0.3cm}
\subfloat[Joint Event and Tweets LDA (ETLDA) \cite{hu2012lda}]{\includegraphics[width= 1.6in]{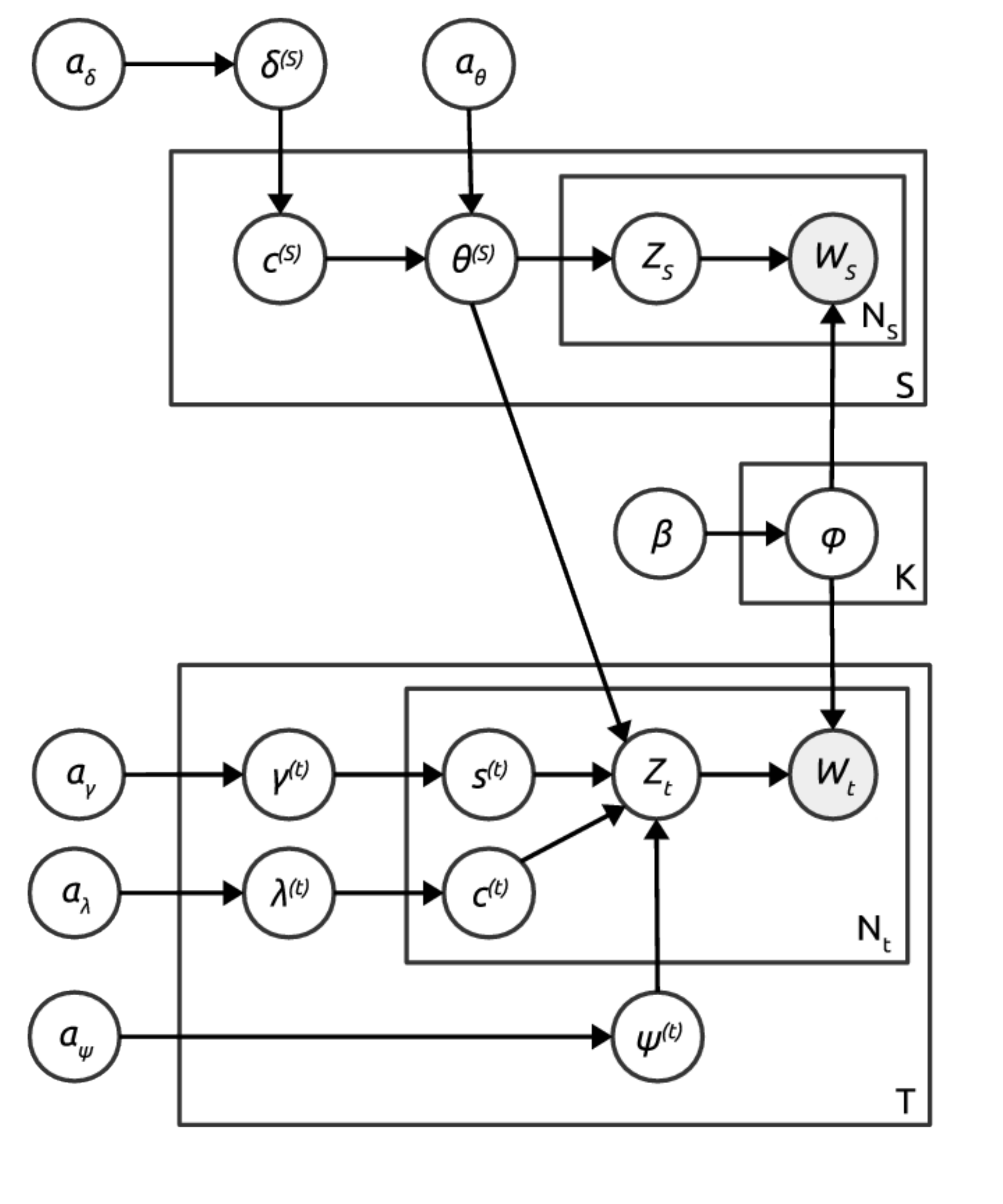}\label{fig:etlda}} \hspace{0.2cm} 
\subfloat[Latent Event and Category Model (LECM) \cite{zhou2015unsupervised}]{\includegraphics[width= 2.1in]{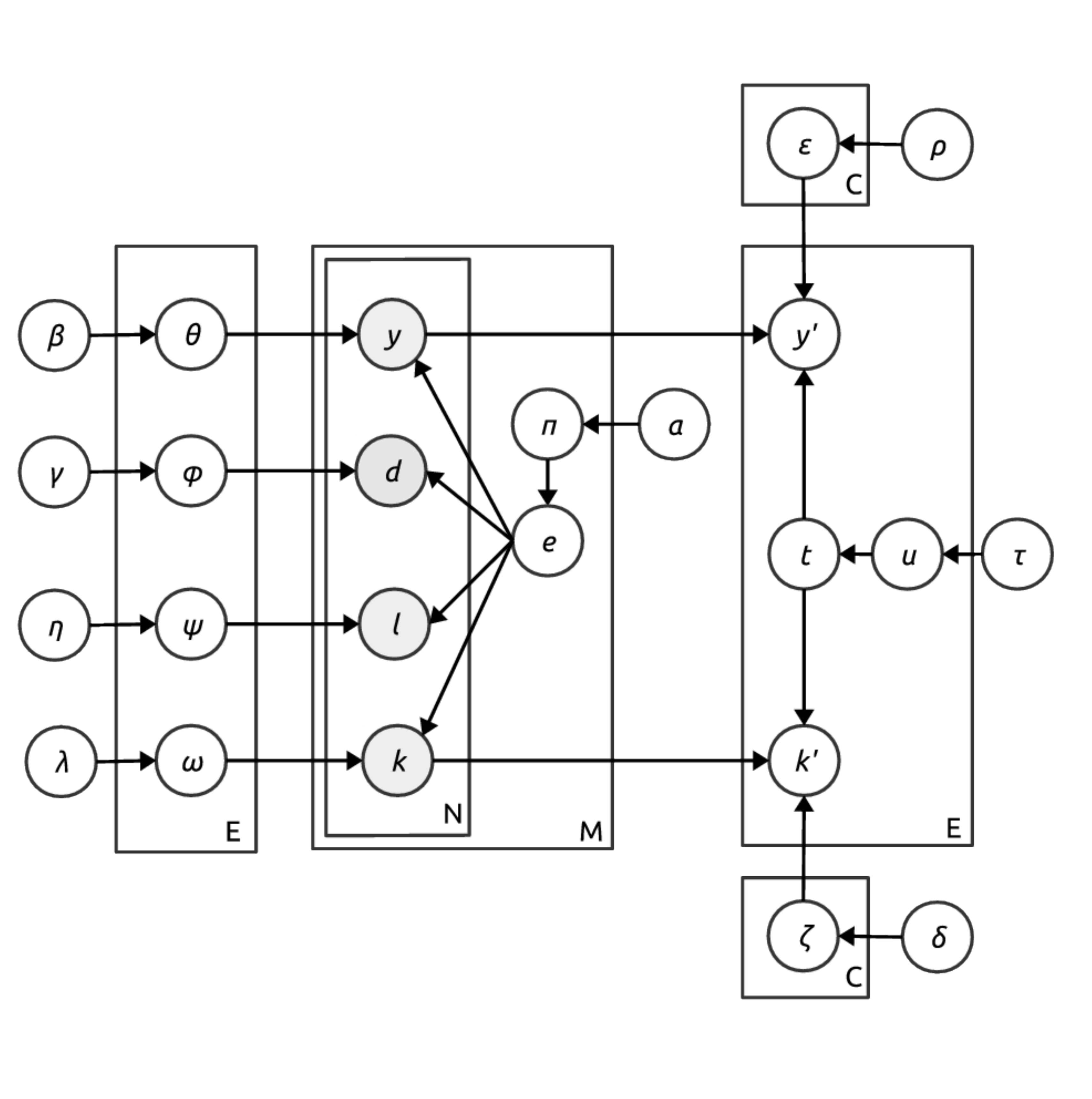}\label{fig:lecm}} \\
\caption{Probabilistic models for event detection, compared to Latent Dirichlet Allocation \cite{blei2002latent}. The parts depicted with dotted lines refer to the original LDA model.}
\label{fig:topic_models}
\end{figure}

This section describes approaches based on probabilistic models that detect events in social media documents in a similar way that topic models identify latent topics in text documents. Originally, topic models relied on word occurrences in text corpora to model latent topics as a mixture of words, and documents as a mixture of the identified set of topics. Latent Dirichlet Allocation (LDA) \cite{blei2002latent}, which is the most known probabilistic topic modeling technique, is a hierarchical Bayesian model where a topic distribution is assumed to have a sparse Dirichlet prior. The model is depicted in Figure \subref*{fig:lda}. $\alpha$ is the parameter of the Dirichlet prior on the per-document topic distribution $\theta$ and $\phi$ is the word distribution for a topic. $K$ denotes the number of topics, $M$ denotes the number of documents while $N$ is the number of words in a document. Given that words $W$ are the only observable variables, the learning of topics, word probabilities per topic, and the topic mixture of each document is tackled as a problem of Bayesian inference solved by Gibbs sampling.

Compared to traditional topic modeling, topic modeling in social media differs mainly because social media documents are multimodal items. Thus, it is not only words in the text that are observable variables, but also other modalities such as the user who posted the document, visual words extracted from embedded images, publication time and quite often the location from where the document originates. In addition, given the short length of documents in social media, and mainly in micro-blogging platforms such as Twitter, the basic assumption that a document is a mixture of topics is more likely to be violated. For that reason many approaches have been proposed that extend traditional models to be aligned with the characteristics of social media data. Some of the most well-known extensions of LDA include TOT \cite{wang2006topics}, TwitterLDA \cite{zhao2011comparing}, and MMLDA \cite{bian2013multimedia}. 

In TOT, depicted in Figure \subref*{fig:tot}, topic discovery is influenced not only by word co-occurrences, but also by the temporal information contained in document timestamps. TwitterLDA, which is depicted in Figure \subref* {fig:twitterlda}, extends standard LDA to model tweets. Each tweet is assumed to be related to a single topic $Z$, which subsequently is influenced by the topic distribution of user $u$ who posted the tweet. In TwitterLDA words in a tweet can be chosen either from a topic-word distribution $\phi$ or from a background distribution  $\phi^B$. Given the fact that social media messages may contain multimedia content, there are topic models that take into account not only  text but also visual content. A notable method in that direction is MultiModal LDA (MMLDA) that uses visual words alongside text words to identify topics. MMLDA is basically an extension of TwitterLDA as shown by Figure \subref*{fig:mmlda}. The same assumptions of a single topic per document, and two word distributions, a topic-specific $\phi^{WG}$ and a background distribution $\phi^{WG}$, are also made. In addition, MMLDA incorporates visual words for each document that is modeled similarly to textual words. To generated visual words, SIFT features are extracted from the images, and a bag-of-visual-words model is applied on them. 

When the target is the detection of events, in contrast to topics, the spatial and temporal dimensions provide important clues. For that reason, approaches aiming at event detection usually employ models built upon the assumptions made by models like TOT, TwitterLDA, and MMLDA. In general, there are two approaches in the use of probabilistic models. The first approach uses these models to represent documents and detect event in a same way as the document-pivot approaches. The second approach includes methods that consider events as latent variables to be learned alongside other variables such as time, location words, and topics. In this section we present methods from both approaches, but we focus mainly on the latter.

Zhou and Chen \cite{zhou2014event} propose a graphical model called Location-Time Constrained Topic (LTT) that extends TOT by incorporating latitude and longitude of social media messages as additional variables $la$ and $lo$ respectively. In other words, the posterior distribution of topics depends not only on words but on three more attributes, time, latitude and longitude, drawn from three Beta distributions $\psi$, $\delta$ and $\gamma$. The plate notation of LTT model is depicted in Figure \subref*{fig:ltt}. Given the LTT model, messages are represented as a distribution over the topics and the similarity between two messages is measured by a distance metric between these distributions based on the Kullback-Leibler (KL) divergence. As the LTT fuses content, time and location attributes, the calculated distance captures an overall content dissimilarity between two messages over these three modalities. The next steps follow the document-pivot paradigm presented in section \ref{sec:doc_pivot}. A complementary measure is proposed that embeds the previous content similarity with a link similarity, based on conversational signals between users. Finally, events are detected using efficient similarity joins over the social media stream of messages.

You et al. \cite{you2013geam} propose an approach for event detection on Twitter that uses a hierarchical Bayesian model named General and Event-related Aspects Model (GEAM). This assumes that in the whole corpus of tweets there are $E$ events and $K$ topics, while each event has $A$ event-related aspects. Figure \subref*{fig:geam} depicts the plate notation of GEAM. Attribute $e$ is a latent variable assigned to each tweet specifying to which event that tweet is related. GEAM also introduces a switch variable $x$, drawn from a binomial distribution that indicates whether a word $W$ in a tweet comes from a general topic word distribution or an event-related aspect word distribution. $\phi$ is a multinomial distribution over words specific to topic $z$ or event $e$, which is drawn from a Dirichlet distribution parameterized by $\beta$. In other words, each word in a given tweet can be associated with a general topic $Z$ or an event aspect $a$ (i.e. time, locations, entities, keywords). Each tweet is modeled not only as a multinomial distribution over topics, as in LDA, but also assigned a latent variable $e$ specifying which event the tweet is describing. The collapsed Gibbs sampling method is used in a similar manner as LDA to estimate the  unknown parameters of GEAM. 

To extract events in tweets and assign events to categories, an unsupervised latent variable model, called Latent Event and Category Model (LECM) was proposed by Zhou et al. \cite{zhou2015unsupervised} and depicted in Figure \subref*{fig:lecm}. To discard noisy tweets, the proposed system trains a lexicon-based binary classifier. Each of the remaining tweets is assigned to a single event $e$, drawn from an event Dirichlet distribution $\pi$, while each event is modeled as a joint distribution over the named entities $y$, date $d$, location $l$ and event-related keywords $k$. The basic assumption in LECM is that tweets containing the same named entities, occurring at the same time and in the same location and having the same keywords are likely to be assigned to the same event. The model also assumes that each event $e$ is assigned to one specific event type $t$, from a set of $C$ event types. That type is modeled as a joint distribution over multinomial distributions $y'$ and $k'$ of semantic classes and event-related keywords respectively. Semantic classes are mapped from named entities using the Freebase API. For example, as described in the paper, a named entity corresponding to a singer could be mapped to the ``music'' class. Compared to GEAM, tweets in LECM are assigned only to an event instance and general topics are missing. 

Cai et al. \cite{cai2015popular} proposed a framework for event detection, tracking and visualization from a collection of Twitter data. For the event detection step, the method uses a model termed Spatio-Temporal Multimodal TwitterLDA (STM-TwitterLDA), and TwitterLDA with MMLDA. The plate notation of STM-TwitterLDA is depicted in Figure \subref*{fig:stm}. Words in tweets are divided into two categories general and topic-specific words as in TwitterLDA and MMLDA. The same holds for the images contained in tweets. However, instead of sampling visual words using Dirichlet multinomial distributions as in MMLDA, STM-TwitterLDA uses CNN features to represent images and model the normalized CNN features by two Dirichlet distributions, a topic-specific and a general one. Two latent random variables $x$ and $y$ act as the general/specific indicators for words and images in a given tweet. In contrast to TwitterLDA that defines a topic distribution per user, in STM-TwitterLDA a topic distribution $\theta$ is shared across tweets collected from the same location. In addition, each topic is a mixture of four distributions: a multinomial distribution over hashtags $\phi$, a multinomial distribution over topic-specific words $\phi^{WS}$, a Beta distribution over timestamps $\psi$, and a Dirichlet distribution over specific images $\beta^{VS}$. There are also two distributions not conditioned on topics: a Dirichlet multinomial distribution over general words $\phi^{WG}$ and a Dirichlet distribution over general images $\beta^{VG}$.
 
Hu et al. propose ET-LDA \cite{hu2012lda}, a joint model used to align events transcripts from an event  with associated Twitter messages. The model is depicted in Figure \subref*{fig:etlda}. It consists of two components, an event and a tweets part, with each having an LDA-like structure and being influenced by a shared topic-word distribution $\phi$. In the event part, the model assumes that an event is formed by sequential segments, each of which discusses a set of topics. A segment consists of one or more paragraphs available from the transcript of the event, with each paragraph $s \in S$ to be associated with a particular distribution of topics $\theta^{(s)}$. The evolution of topics in the event is modeled by associating a binary variable $c^{(s)}$ with each paragraph. Based on the value of this variable, the distribution of topics $\theta^{(s)}$ in a paragraph $s$ is either chosen from a Dirichlet distribution or considered the same as the previous paragraph $s-1$. In the later case, these subsequent paragraphs are merged to form a coherent segment. In the tweets part, the words in a given tweet can be sampled from two distinct topic mixtures depending on a binary variable $c^{(t)}$. Based on the value of $c^{(t)}$ for each word $w$, the word can be drawn either from a distribution of specific topics $\theta^{(s)}$, related to the segments of the event, or from a distribution of general topics $\psi^{(t)}$ over $K$ global topics. In the first case, $\theta^{(s)}$ is associated with a segment $s$ of the event, which is selected according to a categorical distribution $s^{(t)}$. 

Wei et al. \cite{wei2015bayesian} propose a Bayesian graphical model depicted in Figure \subref*{fig:bgm} to discover latent events, identify the spatial and temporal region they occur and associate them with tweets. An event is considered as an extension of topics, having a topical focus but also including a spatial and temporal region in which it occurs. Thus, an event instance $e$, drawn from a multinomial distribution with a prior $\gamma$, is defined as a joint distribution over time, location and words.  Given a set of $E$ possible events a tweet is associated with a single event instance $e$. The location of a tweet associated with event $e$ is assumed to be drawn from a two-dimensional Gaussian distribution $l \sim N(\theta^{(L)}_e, \sigma^{(L)}_e)$, where $\theta^{(L)}_e$ is the geographical center of the event $e$ and $\sigma^{(L)}_e$ its variance. In the same way, the time of a tweet is also modeled as a Gaussian distribution $t$, with mean $\theta^{(T)}_e$ and a variance of $\sigma^{(T)}_e$. The words appearing in a tweet are modeled by a category variable $z$, which is controlled by a multinomial distribution. This variable determines the topic category, therefore the Dirichlet distribution $\phi^{*}$ from which a word $w$ is drawn. $\phi^{0}$ represents global topics that occur across all tweets. $\phi^{L}$ defines a set of region-specific topics. $\phi^{T}$ represents a set of temporally aligned topics that contain words occurring within different temporal factions of the data. $\phi^{T}$ corresponds to topics that are representative of a particular event $e$. Thus, each word $w$ in a tweet is chosen from one of the above topic distributions according to category variable $z$ and the corresponding spatial, temporal and event variables, $l$, $t$ and $e$.

Diao and Jiang \cite{diao2013unified} proposed a model that combines an LDA-like topic model with the Recurrent Chinese Restaurant Process to capture topics and events from a stream of tweets. As in many topic models, topics are modeled as a multinomial distribution over words. Each event is also a multinomial distribution over words. As topics are long-standing and stable, their number is fixed, while for events a non-parametric model, called the Recurrent Chinese Restaurant Process (RCRP), is used to model the birth and death of events. Each user has a multinomial distribution over topics. A latent variable drawn from a user-specific Bernoulli distribution indicates whether a tweet is topic- or event-related. For topic tweets, the topic is sampled from the corresponding user’s topic distribution. For event tweets, the event is sampled according to the Recurrent Chinese Restaurant Process.  

Apart from LDA-inspired approaches, several works were proposed that are based on other probabilistic and statistical models. The paper of Benson et al. \cite{benson2011event} proposes a probabilistic graphical model for the discovery of event records in social media feeds. The approach simultaneously analyzes individual messages (tweets), clusters them according to events, and induces a canonical value for each event property. The graphical model addresses the event detection problem by learning a latent set of records and a record-to-message alignment simultaneously. Records are tuples $<R^1, R^2>$ which represent record’s values for the schema $<ARTIST, VENUE>$. Each message has a sequence labeling, where the labels consist of the record fields and each token in the message has an associated label. Messages, labels and records represented as a probabilistic factor graph and message to record alignment is performed using a variant of the conditional random fields method (CRFs). Although the authors aim at detecting a specific type of event records, the method can be generalized to different and more complex event cases. 

Similarly to the work of Benson et al. \cite{benson2011event}, TwiCal \cite{ritter2012open} provides a structured representation of significant events by extracting a 4-tuple representation of events from Twitter items, which includes a named entity, event phrase, calendar date, and event type. Given a raw stream of tweets, TwiCal extracts named entities, event phrases, and resolved temporal expressions to detect events, which are then categorized to specific types. For named entity recognition an optimized named entity tagger trained on in-domain Twitter data was used. Extraction of event phrases is formulated as a sequence labeling task, solved with Conditional Random Fields for learning and inference. As there are many different ways users can refer to the same calendar date, depending on the publishing date of the tweet, a resolution of temporal expressions is performed to identify when the events described by the event phrases occur. To categorize the extracted events into types an unsupervised approach is proposed based on latent variable models that infers an appropriate set of event types to match the data, and also classifies events into types by leveraging large amounts of unlabeled data. Each event phrase in the data is modeled as a mixture of event types, and each event type corresponds to a distribution over named entities. Inference of that generative model is similar to an LDA variant, termed LinkLDA \cite{erosheva2004mixed}. Finally, extracted events are ranked by measuring the strength of the association between an entity and a specific date in the event. The strength is estimated by the $G^2$ log-likelihood ratio statistic between  the occurrences of the entity and the date in the tweets of the dataset.

\section{Event-based Retrieval and Summarization}
\label{sec:retrieval_summarization_methods}

Given real-world events, which are detected automatically or can be introduced in the form of user queries, there is a need for methods that identify and organize additional social media content associated with these events. When events take the form of user provided queries, there is no content available. In case of automatically detected events, the content used for the detection is usually associated with the events, but additional content that augments the event, can be discovered. In both cases, the discovery of relevant content can be seen an information retrieval problem. This task is different from the traditional document search problem, where a user's information need is fulfilled by one or more relevant documents. In social media applications, information needs cannot be satisfied by a limited set of social media posts due to the limited content and context provided by individual messages. 
To add substantial value for users, more structured and diverse lists must be returned. Furthermore, due to the increasing popularity of micro-blogging platforms, the amount of event-related posts has reached impressive levels. Also, event-related streams or collections are highly diverse, with different associated topics and conversations among users, and a high degree of redundancy. In addition, as relevance and significance of the messages may vary, there is a profound need for event-based summarization mechanisms that can produce concise summaries, covering the main aspects of the event. 

\subsection{Event-based retrieval}
\label{sec:event_based_retrieval}

An information need is usually expressed as a simple and short query, consisting of few terms. That makes the finding of relevant messages a challenging task, since social media messages may not contain any of the terms even if they are related to the query. Furthermore, when it comes to event-related queries, the events have a temporal and very often a spatial dimension that should be considered in the retrieval process. To this end several methods have been proposed to support information retrieval on social media messages and mainly on micro-blog collections \cite{choi2012temporal,choi2012quality,miyanishi2013improving,severyn2015learning}. For the evaluation of retrieval approaches in micro-blogs a special track\footnote{\url{http://trec.nist.gov/data/microblog.html}} took place as part of the Text REtrieval Conference (TREC) from 2011 to 2015. Given a set of topics of interest, expressed by title, description and time, the scope of the track was the retrieval of relevant content from collections of tweets. Although, not limited to event-based retrieval, these micro-blog retrieval approaches are the basis for event-oriented retrieval. Additionally, efficient and accurate retrieval is an important step of the event summarization process, as relevance calculation is a core part of summarization. 
 
\begin{figure}
\centering
\includegraphics[width=0.9\textwidth,trim={0 1cm 0 1cm},clip]{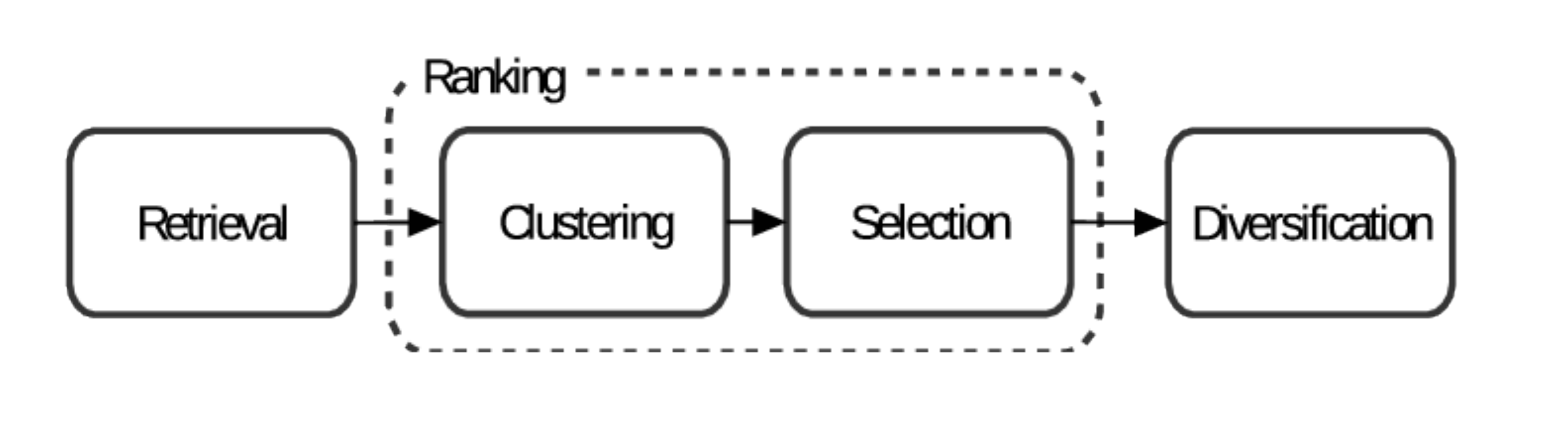}
\caption{A general processing pipeline for event summarization.}
\label{fig:event_summarization_pipeline}
\end{figure}

Becker et al. \cite{becker2011automatic} proposed a two-step system to obtain Twitter messages related to planned events. Given an event, the first step is to build a query derived from the event, aiming at high precision results. Queries are formed based on the structured description of an event, and its other aspects, such as the venue the event takes place. Intuitively, these highly restrictive queries return only messages that relate to the planned event, but at the expense of recall. To improve recall, term-frequency analysis is performed on the high-precision tweets to identify descriptive terms and phrases, which are used, in turn, to define new queries. A rule-based classifier is then used to select among this new candidate queries, and then use them to retrieve additional event messages. 

Becker et al. \cite{becker2012identifying} extended their previous work \cite{becker2011automatic} to support identification of content for planned events across multiple social media sites. To this end, the authors developed a set of query formulation strategies used to retrieve content associated with an event. The main motivation is that some platforms like Twitter can be useful as a mean for timely event detection, while others like Flickr or YouTube can be used to augment the events with additional multimedia content. Similar to \cite{becker2011automatic}, a two step query-generation approach is proposed for the retrieval of content given an event. First a precision-driven set of queries is constructed based on event attributes such as title, description, date and location. These queries are directed to each target platform to retrieve a limited but highly relevant set of  social media documents. Two approaches are then proposed to formulate new recall-driven queries based on these documents. The first is to extract the most frequently used $n$-grams, while weighing down terms that are naturally common in the English language. To normalize $n$-grams frequency, a language model built from web documents was used. The second approach aims to identify meaningful event-related concepts in the ground-truth data using an external reference corpus. A web-based term extractor was used that leverages a large collection of web documents and query logs to construct an entity dictionary, and uses it along with statistical and linguistic analysis methodologies to find a list of significant terms. In addition, the use of event content from one social media site to help retrieve event documents from another social media site is investigated. This is useful when the precision-oriented strategies do not retrieve results from all sites. For example, if the precision-oriented queries fail to retrieve any documents for one site, and hence the method cannot generate any recall-oriented query for that site, documents retrieved from another platform can be used as input to the query construction step. 

Liu et al. \cite{liu2011finding} presented a method that combines semantic inferencing and visual analysis aiming at finding media (photos and videos) illustrating events in an automatic way. As a first step in the procedure, given some event descriptions, event directories such as \textit{Last.fm}, \textit{Eventful} and \textit{Upcoming} are used to obtain structured information about the events, which are then converted into the LODE ontology \cite{shaw2009lode}. Explicit relationships between scheduled events and photos hosted on Flickr can be looked up using special machine tags. However, the set of images available on the web that can be explicitly associated to events using machine tags is generally a tiny subset of all media that are actually relevant to the event. To this end, the authors investigate several approaches to find these images, querying by title, geo-location and time and  pruning irrelevant media. 

Metzler et al. \cite{metzler2012structured} tackled the problem of structured retrieval of historical event information over micro-blog archives, by proposing a novel temporal query expansion technique. Given a user query, the approach first retrieves a set of timespans for which the query keywords were most heavily discussed. Timespans are subsequently ranked according to the proportion of messages posted during them. Considering these timespans as related to the event is a kind of pseudo-relevance feedback mechanism. 
For each pseudo-relevant timespan, a burstiness score is computed for all terms that occur in the messages posted during it. The idea is that during a timespan in which a query is being heavily discussed another term that is trending might also be related to the event. The final step of the query expansion process is to aggregate the burstiness scores across all pseudo-relevant timespans to generate an overall score for each term. The $k$ highest weighted terms are then used for expansion. The expanded query is then used to retrieve additional content and identify the highest scoring timespans in the same way as in the first step. The final step of the retrieval process is to produce a short summary for each retrieved timespan. This not only returns ranked individual messages, but also a ranked list of historical event summaries, with respect to the original query. Given a timespan, the summary is produced by a simple IR language model that selects messages that are found to be the most relevant to the expanded representation of the original query in that timespan.  

\subsection{Event-based summarization}
\label{sec:event_based_summarization}

Given a set of event-related messages, for example by using an approach such as the previous ones, 
there is often a need to select only a representative subset of these. A substantial body of research efforts attempt to solve this text summarization problem, which is a special case of the Multi-Document Summarization (MDS) problem. One of the first MDS approaches relies on the generation of a content-based centroid. Then, the summary of a set of documents, represented by TF-IDF vectors, consists of those documents that are closest to the centroid. Graph based approaches have also been proposed to detect salient sentences from multiple documents, with LexRank \cite{erkan2004lexrank} being the most notable among them. This first constructs a graph of sentences (nodes), with the textual similarity between two sentences serving as the connection (edge) between them. Then, it computes the saliency of each sentence using some centrality measure, such as the Eigenvector Centrality or its well known variant, PageRank. These approaches are based on the observation that the most central messages in a cluster are more likely to reflect key aspects of the event than other, less central messages. One of the first works that use centrality-based methods to select subsets of messages is that of Becker et al. \cite{becker2012identifying}, which propose a centrality-based approaches to extract high-quality and relevant tweets with respect to a specific event. 

However, the text brevity, informal writing and non-grammatical character of many micro-blogging posts, as well as the diversity of the underlying topics make the summarization problem much more challenging in the context of social media compared to the traditional MDS setting, where the input collection consists of long well-formed documents. In addition, the temporal dimension, which is a crucial element of micro-blogging posts, and the social interaction between users in social media platforms, are disregarded by previous MDS methods. To this end, numerous methods were proposed that incorporate not only the textual information of documents, but also their temporal and social aspects. The core idea of the majority of previous works is the clustering of documents set into coherent topics or sub-events and the selection of the most ``representative'' documents in each segment. Although there are works that investigate the use of social dimension to the problem of event detection \cite{guille2014mention}, to our knowledge, this dimension is usually disregarded, compared to the number of methods that are based on content and temporal information.

Nichols et al. \cite{nichols2012summarizing} describe a sports event summarization algorithm. Given a collection of tweets related to a specific event, the method uses a peak detection algorithm to detect important moments in the timeline of tweets. Then, it applies a graph-based technique to extract important sentences from the tweets around these moments. In \cite{chua2013automatic}, the authors propose a probabilistic model for topic detection in Twitter that handles the short length of tweets and considers time as well. Instead of relying only on the co-occurrences of words (as the majority of traditional probabilistic text models do), the proposed model uses the temporal proximity of posts to reduce the sparsity of the term co-occurrence matrix. Then, for each detected topic, the method considers the set of tweets with the highest similarity to the topic word distribution as the most representative. Shen et al. \cite{shen2013participant} present a participant-based approach for event summarization, which first detects the participants of the event, then applies a mixture model to detect sub-events at participant level, and finally selects a tweet for each detected sub-event based on the TF-IDF centroid approach. In a similar work, Chakrabarti and Punera \cite{chakrabarti2011event} propose the use of a Hidden Markov Model to obtain a time-based segmentation of the stream that captures the underlying sub-events.

Recent works have focused on the creation of visual event summaries based on messages and content shared on social media. TwitInfo \cite{marcus2011twitinfo} is a system for summarizing events on Twitter through a timeline display that highlights peaks of high activity. Alonso and Shiells \cite{alonso2013timelines} create football match timelines annotated with the key match aspects in the form of popular tags and keywords. Dork et al. \cite{dork2010visual} propose an interface for large events employing several visualizations, e.g., image and tag clouds. However, the aforementioned methods only make use of textual and social features for creating visualizations, and ignore the visual content of the embedded multimedia items.

The increasing use of multimedia content in micro-blog platforms has motivated numerous studies to consider visual information alongside text. Bian et al. \cite{bian2013multimedia} proposed the use of MMLDA (cf. section \ref{sec:topic_models}) for detecting topics simultaneously taking into account the textual and visual content of posts with embedded images. The topics identified by MMLDA are used to calculate coverage and diversity for each image in the dataset. Finally, the output of this method is a set of representative images for the underlying event that maximize these measures. A slightly different problem was tackled by Lin et al. \cite{lin2012generating}. Unlike other methods that generate summaries as sets of posts or images, this method aims to create a storyline from a set of event-related multimedia objects. To this end, it constructs a multi-view graph of objects, with two types of edges, visual and textual, capturing the content similarity along with the temporal proximity among objects. Then, it extracts a time-ordered sequence of important objects based on Steiner trees \cite{wang2012generating}. 

Meladianos et al. \cite{meladianos2015degeneracy} proposed a system that, given an event-related set of tweets, identifies the key moments (sub-events) of the event, and select a tweet to best describe each sub-event using a simple yet effective heuristic. The input set of tweets is represented as a graph-of-words, where nodes correspond to words extracted from the whole set, while edges indicate that the two terms co-occur in any tweet of the input set. Then, using the graph degeneracy concept, term weights are extracted. To detect sub-events, the stream is divided in time intervals for which a graph-of-words is generated. Then, using the weights of the terms as determined by the degeneracy-based weighting mechanism, a sub-event is reported if there is a sharp increase in the terms, regardless of the number of tweets in that time-interval. If a sub-event has been detected, then the tweets in that time interval are scored, by the the sum of their term weights, and the highest-scored tweet is selected.

Gao et al. \cite{gao2012joint} propose a joint topic modeling method to summarize events across news and social media streams. The method leverages a topic modeling approach, termed cross-collection Topic-Aspect Model (ccTAM), that combines cross-collection LDA (ccLDA) \cite{paul2009cross} with a Topic-Aspect Model (TAM) \cite{paul2010two}. ccLDA is an extension of LDA, in which each topic is associated with two classes of word distributions: a shared distribution among all collections, and one that is specific to the collection from which the document comes. TAM decomposes each document not only into a mixture of topics, but uses a second mixture that represents the underlying aspects in the collection of documents such as different perspectives and viewpoints. The novelty in the work of Gao et al. is that the proposed ccTAM uses two complementary types of information as input, news and tweets, which are represented in a joint fashion as mixtures of topics and aspects. To produce a summary of the event, a bipartite-graph ranking algorithm is used, where the nodes at one side correspond to sentences extracted from the news and those at the other side correspond to tweets. As a transition probability is necessary for the edge weights between tweets and sentences, the output ccTAM is used to calculate a complementary measure between these two parts. Finally, the top ranked sentences and tweets at both sides are used to generate a news summary and a tweet summary in such a way that the predefined length of the summaries are met.

MacParlane et al. \cite{mcparlane2014picture} proposed a method to select and rank a diverse set of images with a high degree of relevance to a 
given event. A unique characteristic of this work is the use of external sources of multimedia content, such as websites, in cases where the amount of images in the posts of an event is insufficient for the creation of meaningful visual summaries. To detect and remove near-duplicates among the images, and thus to increase visual diversity, the Perceptual Hash (pHash) method \footnote{\url{http://www.phash.org/}} is used to cluster images into sets of near-duplicates, such as as brightness or contrast adjusted images, resized or slightly rotated, or images with embedded watermark. Then, they use visual features in conjunction with an SVM classifier to discard irrelevant images and images of low quality, such as memes and images with heavy text. Finally, they evaluate numerous ranking methods based on clustering, for selecting a small number of representative images for the event. 

Schinas et al. \cite{schinas_icmr2015} proposed a framework that uses an event-related set of social media items, to create a visual summary that describes the main moments of the event. To overcome the problem of spam and noisy messages, the framework first discards irrelevant and low quality messages, using a set of filters based on textual and visual clues. Then a multi-graph is generated, in which vertices correspond to messages, while four types of edge represent different types of association between the messages. More specifically, two types of undirected edges are used to represent textual and visual similarity, and two types of directed edges represent temporal proximity and social interactions (e.g. shares, replies, etc). Since there is a high degree of redundancy in social media collections, 
a visual de-duplication step follows. Using the subgraph corresponding to messages with embedded multimedia, the Clique Percolation Method is used to find and merge sets of messages that are visual duplicates. Then, to detect topics and sub-events on a main event, the Structural Clustering Algorithm for Networks (SCAN) is used. To create a visual summary, the messages are ranked according to a score that integrates social attention topic coverage. Finally, messages having images are re-ranked using the DivRank iterative scheme, a variant of PageRank, designed for improving content diversity in the ranking process.   

\section{Datasets and Evaluation}
\label{sec:datasets_evaluation}

To evaluate the performance of event detection methods several metrics have been proposed and used in the related literature, with precision and recall being the most common among them. Given a set of candidate events reported by a method, precision measures the percentage that correspond to actual events. Recall on the other hand, indicates the percentage of events that the methods managed to detect out of the whole set of reference events in the dataset. However,these metrics alone are not sufficient, as they cannot capture the quality of the detected events. This is more evident in the case of document-pivot and topic modeling approaches, where an event is reported as a set of documents related to the event. Thus, there is a need for metrics that compare detected and actual events at the level of their documents.  

Given a set of events and a dataset of related documents combined with non-event documents, the performance of a method that returns documents for each of the events can be measured by calculating precision and recall at the level of documents. Precision is the fraction of returned documents that truly belong to an event, while recall measures the fraction of all event-related documents returned by an approach. Macro F-score is the harmonic mean of precision and recall. Thus, it measures the goodness of the retrieved photos, but not the number of retrieved events, or how accurate the correspondence between retrieved images and events is. This is done by the Normalized Mutual Information (NMI)  \cite{strehl2002cluster}, which compares two sets of  clusters, where each cluster comprises the documents of a single event), jointly considering the goodness of the retrieved documents and their assignment to different events.
Another metric is B-Cubed \cite{amigo2009comparison} that estimates the precision $P_b$ and recall $R_b$ associated with each document in the dataset individually, and then uses the average precision and recall values for the dataset to compute their harmonic mean: 

\[
	P_b = \sum_{d \in D} \frac{1}{|D|} \frac{|\textrm{Cluster}(d) \cap \textrm{GroundTruth}(d)| }{|\textrm{Cluster}(d)|}
\]
    
\[    
    R_b = \sum_{d \in D} \frac{1}{|D|} \frac{|\textrm{Cluster}(d) \cap \textrm{GroundTruth}(d)| }{|\textrm{GroundTruth}(d)|}
\]

\[    
	\textrm{B-Cubed} = 2 \cdot \frac{P_b \cdot R_b}{P_b + R_b}
\]
To evaluate event-based retrieval and summarization methods, different evaluation metrics have been proposed. Regarding the retrieval of content, the well known Precision and Recall are used. However, in the context of event summarization, more sophisticated metrics are needed that take into account the coverage of different aspects of the event, the redundancy in the retrieved set of items, the diversity in text and visual content, etc. Thus in the works presented in section \ref{sec:retrieval_summarization_methods}, a wide set of metrics have been proposed and used. Typically, to evaluate summarization results, a golden standard is used that represents an ideal summary. In social media data, this golden standard usually takes the form of a set of items that summarize the dataset. Given these items, Precision, Recall and F-measure can be calculated between the produced summary and the golden standard. However, this may be misleading as the produced summary may contain not the exact set of items, but similar items that convey the same information. For that reason the so-called ROUGE \cite{lin2003automatic} metrics are used, a family of metrics that measure the similarity between an automated summary and a set of manual summaries, using n-gram co-occurrence statistics between the sets. The simplest among them is the ROUGE\textsubscript{N} metric:
\[
	ROUGE_N = \frac {\sum_{s \in MS}{\sum_{ng \in S} \textrm{Match}(ng)}} 
    {\sum_{s \in MS}{\sum_{ng \in S} \textrm{Count}(ng)}}
\]
where $\textrm{Count}(ng)$ is the number of n-grams in the manual summary, and $\textrm{Match}(ng)$ is the number of co-occurring n-grams between the manual and automated summaries.

McParlane et al, whose work \cite{mcparlane2014picture} was presented in section \ref{sec:retrieval_summarization_methods} proposed the use of three precision-based and two diversity-oriented metrics to evaluate visual summaries produces by their approach: 
\begin{itemize}
\item \textbf{Precision ($P@N$):} The percentage of relevant images among the top $N$ images in the summary.
\item \textbf{Success ($S@N$):} Given summaries for multiple events, the percentage of events, for which there exists at least one relevant image among the top $N$ returned by the method.
\item \textbf{Mean Reciprocal Rank (MRR):} Computed as $1/r$, where $r$ is the rank of the first relevant image returned, averaged over all summaries.
\item  \textbf{$\alpha$-Normalised Discounted Cumulative Gain ($\alpha$-$nDCG@N$):} This metric computes the usefulness, or gain, of an image based on its position in the ranked list. Parameter $\alpha$ balances the importance of relevance and diversity.
\item \textbf{Intent-aware Expected Reciprocal Rank (ERR-$IA@N$):} For this metric, the contribution of each image is based on the relevance of images ranked above it, by computing the ERR for each sub-event, with a weighted average computed over sub-events.
\end{itemize}
Despite the proliferation of research methods for event detection and retrieval, the available public datasets for their evaluation are limited. For that reason, most of the works presented in section \ref{sec:event_detection} are evaluated on different datasets collected and annotated by their authors. Given the lack of a commonly accepted ground truth dataset, containing annotations on the actual events, most of the works follow a qualitative analysis of the detected events. Also, they may report the precision of the detected events by examining which of them correspond to actual events. However, it is difficult to calculate the recall as the total set of actual events that are present in a collection are unknown. In addition, judgments about the association of documents to events are also missing, so it is difficult to evaluate the obtained results, especially those produced by document-pivot methods that provide events as sets of documents. As a result, an explicit comparison between these approaches is difficult or even impossible to perform. 

Table \ref{tab:event_detection_results} presents the reported results for selected event detection methods. The listed papers have conducted experiments on proprietary datasets for which there is no ground truth regarding the association of social media posts to events, but rather only a list of known events. For that reason, the papers report the number of candidate events detected by the methods and the precision achieved at the event level (i.e. $x\%$ of the detected events are correct). This is the case for instance for methods such as the ones presented in \cite{becker2011beyond,guille2014mention,petrovic2010streaming} that report precision only for the highest ranked clusters (events). However, the highest ranked clusters are indeed more likley to be related to real-world events. For that reason, a fixed number of clusters $K$ is considered during evaluation and Precision@$K$ ($P@K$) is calculated. Still, this kind of evaluation does not take into account the quality of clusterings, i.e. how accurately the social media items are assigned to clusters. 
An additional set of reported results are presented in Table \ref{tab:multimodal_event_detection_results} with an emphasis on event detection from multimodal items (Flickr images). It is interesting that despite the different datasets that are used by different researchers the reported NMI scores lie in the same range ($\sim 0.95$).

\begin{table}
\centering
\caption{Reported results for selected event detection methods in Twitter. In the cases that the number of reported events is marked with *, the number is selected by the authors to report performance in terms of $P@K$.}
\begin{tabular}{|l|c|c|c|c|}
\hline
\multirow{2}{*}{\textbf{Work}} & \multirow{2}{*}{\textbf{Tweets}} & \multicolumn{3}{c|}{\textbf{Results}} \\ \cline{3-5} 
 &  & \textbf{Method} & \textbf{Events} & \textbf{Precision} \\ \hline
Weng \& Lee, 2011 \cite{weng2011event} & 4,331,937 & EDCoW  & 21 & 76.2\% \\ \hline
\multirow{2}{*}{Li et al., 2012 \cite{li2012twevent}} & \multirow{2}{*}{Same as \cite{weng2011event}} & EDCoW & 21 & 76.2\% \\ \cline{3-5} 
 &  & Twevent & 101 & 86.1\%  \\ \hline
Parikh \& Karlapalem, 2013 \cite{parikh2013events} & 1,023,077 & ET  & 23 & 91\% \\ \hline
\multirow{4}{*}{Guille \& Favre, 2014 \cite{guille2014mention}} & \multirow{2}{*}{1,437,126} & ET & 40* & 57.5\% \\ \cline{3-5} 
 & & MABED & 40* & 77.5\%  \\ \cline{2-5} 
 & \multirow{2}{*}{2,086,136} & ET  & 40* & 70\% \\ \cline{3-5} 
 & & MABED & 40* & 82.5\% \\ \hline
Becker et al., 2011 \cite{becker2011beyond} & 2.6M & RW-Event & 5*, 10* & 85\%, 70\% \\ \hline
Petrovi\'{c} et al., 2010 \cite{petrovic2010streaming} & 163.5M  & LSH & 1000* & 34.0\% \\ \hline
\end{tabular}
\label{tab:event_detection_results}
\end{table}

\begin{table}
\centering
\caption{Evaluation of multimodal event detection from Flickr images.}
\begin{tabular}{|l|c|c|c|c|} \hline
\multirow{2}{*}{\textbf{Work}}   & \multirow{2}{*}{\textbf{Images}} & \multirow{2}{*}{\textbf{Events}} & \multicolumn{2}{c|}{\textbf{Results}}                \\ \cline{4-5} 
                                                                 &                                  &                                  & \multicolumn{1}{c|}{\textbf{NMI}} & \textbf{B-Cubed} \\ \hline
Becker et al., 2009 \cite{becker2009event}                                    & 270,425                          & 9,515                            & 0.933                             & -                \\ \hline
\multirow{2}{*}{Becker et al., 2010 \cite{becker2010learning}} & 270,425                          & 9,515                            & 0.951                            & 0.826           \\ \cline{2-5} 
                                                                 & 594,946                          & 24,958                           & 0.944                            & 0.816           \\ \hline
Reuter \& Cimiano, 2012 \cite{reuter2012event}                 & $\sim$1M                  & 36,782                           & -                                 & 0.744            \\ \hline
Petkos et al., 2012 \cite{petkos2012social}                    & 73,645                           & 36                               & 0.915                             & -                \\ \hline
\multirow{2}{*}{Petkos et al., 2017 \cite{petkos_mtap2017}}   & 306,159                          & 14,852                           & 0.975                             & -                \\ \cline{2-5} 
                                                                 & 362,578                          & 17,834                           & 0.981                             & -                \\ \hline
Bao et al., 2013 \cite{bao2013social}                          & 167,332                          & 149                              & 0.708                             & -                \\ \hline
\end{tabular}
\label{tab:multimodal_event_detection_results}
\end{table}

One of the few publicly available benchmarks is the Social Event Detection (SED) task, an open challenge that ran within the MediaEval Initiative\footnote{\url{http://www.multimediaeval.org/}} for four successive years (2011-2014). The main objective of the task was the discovery of social events in collections of multimedia content posted in social media platforms. The SED 2011 task \cite{papadopoulos2011social} had two challenges. In both challenges participants were provided with a set of Flickr images and were asked to retrieve events of a particular type at particular locations. The first challenge sought for all soccer events in Barcelona (Spain) and Rome (Italy), while the second challenge required from the participants to find all events that took place in May 2009 in the venue named Paradiso in Amsterdam and in the Parc del Forum in Barcelona. Note that the actual events were unknown to  participants, who had to provide the events as sets of images related to them. The dataset for the 2011 task consisted of 73,645 Flickr photos. 

The SED 2012 task \cite{papadopoulos2012social} included three challenges, quite similar to those of the previous year. Each challenge required the identification of events that met some predefined criteria: the first challenge required participants to find images associated with technical community events taking place in Germany; the second asked participants to find all  soccer events taking place in Hamburg (Germany) and Madrid (Spain); finally, in the third challenge, the target was demonstrations and protest events of the ``Indignados movement'' occurring in public places in Madrid. A much wider collection of 167,332 photos by 4,422 unique Flickr users was created\footnote{\url{http://mklab.iti.gr/project/sed2012}}, and the collected photos were all licensed under a Creative Commons licence, and were captured between the beginning of 2009 and the end of 2011. 

The 2013 SED task \cite{reuter2013social} had a significant difference to the two previous years’ tasks. Whereas in the previous years, a single dataset that includes both event and non-event photos was provided and the challenges asked for the retrieval of content about specific events, in 2013 two datasets were provided, and two new distinct challenges were defined. The first challenge required participants to provide a complete clustering of images in the dataset according to events. That is, the first challenge asked for a clustering of all images in the relevant dataset, according to the events that they depict. This comes in contrast to the challenges in the first two years, where a) not all images in the collection were related to some event and b) specific criteria were defined for the events of interest. The second challenge required the classification of images into specific event types. The dataset\footnote{\url{http://mklab.iti.gr/project/sed2013}} for the first challenge consists of 427,370 pictures from Flickr and 1,327 videos from YouTube together with their associated metadata. The pictures, uploaded between January 2006 and December 2012, correspond to 21,169 events. The dataset for Challenge 2 was collected from Instagram. The training set was collected between 27th and 29th of April 2013, based on event-related keywords, and consisted of 27,754 pictures, while the test set was collected between the 7th and 13th of May 2013 and consisted of 29,411 pictures.

The 2014 SED task \cite{petkos2014social} consisted again of two challenges/subtasks. For the first challenge similar to 2013, participants were asked to produce a full clustering of the images, so that each cluster corresponds to a social event. In the second subtask, a collection of events was provided; each event was represented by a set of images with their metadata, and participants were asked to retrieve those events, i.e. other images, that met some criteria. The test dataset\footnote{\url{http://mklab.iti.gr/project/sed2014}} contains 362,578 images and together with it, the grouping of these images into 17,834 social events. The training dataset, for which grouping of images into clusters is provided, contained 110,541 images.

As there was a difference across the challenges from year to year, for their evaluation different approaches were tested. For the challenges in the first two years, 2011 and 2012, two metrics were used, Macro F-measure and NMI. For the first challenge in years 2013 and 2014, which can be seen as a full clustering problem, NMI was also used, in conjunction with the micro F-measure (B-Cubed). Tables \ref{tab:sed_2013_results} and \ref{tab:sed_2014_results} present an overview of the reported results of SED approaches in the task editions for 2013 and 2014 respectively. Regarding the first challenge in both years, most of the participants achieved a decent performance and in many cases the results were surprisingly good. The second challenge of 2013, targeting event type classification, led to results that were lower than expected. Participation in the 2014 second challenge was limited, and the mediocre reported results indicate that there is space for improvements in the event-based retrieval task. 


\begin{table}
\centering
\caption{SED 2013 results}
\begin{tabular}{|l|cc|cc|} \hline
 & \multicolumn{2}{|c|}{Challenge 1} & \multicolumn{2}{|c|}{Challenge 2} \\ \cline{2-5}
 & F-score     & NMI & F\textsubscript{cat} & F\textsubscript{E/NE}\\ \hline
Rafailidis et al., 2013 \cite{rafailidis2013data} 			& 0.570	&  0.873 & - 		& - 	\\
Samangooei et al, 2013 \cite{samangooei2013social} 		& 0.946	&  0.985 & - 		& - 	\\
Schinas et al., 2013 \cite{schinas2013mediaeval}			& 0.704	&  0.910 & 0.334 	& 0.716 \\
Manchon-Vizuete \& Gir\'{o} Nieto, 2013 \cite{manchon2013upc} 				& 0.883	&  0.973 & - 		& - 	\\
Nguyen et al., 2013 \cite{nguyen2013event}  			& 0.932	&  0.984 & 0.449 	& 0.854 \\
Zeppelzauer et al., 2013 \cite{zeppelzauer2013unsupervised}  & 0.780	&  0.940 & - 		& - 	\\
Sutanto \& Nayak, 2013 \cite{sutanto2013admrg}  			& 0.812	&  0.954 & 0.131 	& 0.537 \\
Wistuba \& Schmidt-Thieme, 2013 \cite{wistuba2013supervised}  		& 0.878	&  0.965 & - 		& - 	\\
Tserpes et al., 2013 \cite{tserpessimilarity}  			& 0.236	&  0.664 & - 		& - 	\\
Gautam et al., 2013 \cite{itikavit}  					& 0.142	&  0.180 & - 		& - 	\\
Brenner \& Izquierdo, 2013 \cite{brenner2013mediaeval}  		& 0.780	&  0.940 & 0.332 	& 0.721 \\ \hline
\end{tabular}
\label{tab:sed_2013_results}
\end{table}

\begin{table}
\centering
\caption{SED 2014 results}
\begin{tabular}{|l|cc|ccc|} \hline
 & \multicolumn{2}{|c|}{Challenge 1} & \multicolumn{3}{|c|}{Challenge 2} \\ \cline{2-6}
 & F-score & NMI & P & R & F\textsubscript{1} \\ \hline
Zaharieva et al., 2014 \cite{zaharieva2014clustering}	& 0.948 	& 0.989 & 0.420 & 0.406 & 0.288	\\
Guinaudeau et al., 2014 \cite{guinaudeau2014limsi}		& 0.821 	& 0.955 & - & - & - 		\\
Riga et al., 2014 \cite{riga2014certh}			& 0.916 	& 0.982 & 0.708 & 0.392 & 0.460 \\
Manchon-Vizuete et al., 2014 \cite{manchon2014upc}			& 0.924	    & 0.982 & - & - & - 		\\
Sutanto \& Nayak, 2014 \cite{sutanto2014ranking}		& 0.753 	& 0.902 & - & - & - 		\\
Denman et al., 2014 \cite{denman2014saivt}			& 0.753 	& 0.902 & - & - & - 		\\ \hline
\end{tabular}
\label{tab:sed_2014_results}
\end{table}

Furthermore, McMinn et al. \cite{mcminn2013building} proposed a methodology for building a corpus of tweets, to be used for event detection methods. To create a candidate list of events, McMin et al. used a number of existing event detection approaches and the Wikipedia Current Events Portal\footnote{\url{https://en.wikipedia.org/wiki/Portal:Current_events}}. The two methods used to detect events were the ones by Petrovi\'{c} et al. \cite{petrovic2010streaming} and Aggarwal \& Subbian \cite{aggarwal2012event}, both presented in section \ref{sec:feature_pivot}. The authors ran the first method \cite{petrovic2010streaming} over a random sample of tweets using parameters similar to those reported in the original paper. However, rather than ranking clusters every 100,000 tweets, authors opted for a ranking at an hourly basis. For each hour, clusters were ranked by the number of unique users, and clusters with low entropy were removed. This yielded a list 1,340 candidate events. Using the cluster summarization method of Aggarwal \& Subbian \cite{aggarwal2012event}, McMinn et al. were able to produce 1,097 candidate events. Wikipedia events were used within an information retrieval framework to obtain tweets related to the events. Each Wikipedia event was represented as a short description, a category, and a link to a relevant news article. The event description was then used as an initial query to retrieve tweets that potentially discuss the event. Query expansion was also performed using some of the best performing approaches in the TREC\footnote{\url{http://trec.nist.gov/}} Microblog track \cite{amati2011fub,hoang2012ugent}. For each of the 468 events on the Wikipedia Current Events Portal listed between the dates covered by the corpus, the top 2,000 tweets were retrieved from a window of 72 hours, centered around the date of the event. A by-product of this work is a corpus of about 500 news events sampled over a four-week period, including relevance judgments for thousands of tweets referring to them.

For evaluation of event summarization, McParlane et al. \cite{mcparlane2014picture} created a dataset, built upon the one by McMinn et al. \cite{mcminn2013building}. This dataset is available on demand. Another public dataset, used for the evaluation of the approach proposed in \cite{schinas_icmr2015} and \cite{schinas_ijmir2016} is also available\footnote{\url{https://github.com/MKLab-ITI/mgraph-summarization}}. This dataset consists of an extension of McParlane's dataset\cite{mcparlane2014picture}, alongside two more datasets related to the 2015 Baltimore protests\footnote{\url{https://en.wikipedia.org/wiki/2015_Baltimore_protests}} and the 2012 US Presidential Elections\footnote{\url{https://en.wikipedia.org/wiki/United_States_presidential_election,_2012}}.

The TREC Real-Time Summarization Track\footnote{\url{http://trecrts.github.io/TREC2017-RTS-guidelines.html}} is the successor of the Microblog Track since 2016. The task focuses on the problem of topic-oriented summarization of a Twitter stream. Given a set of topics of interest, the participants should identify relevant tweets in the stream per topic, and update a summary for each topic when new pieces of information are presented. There are two scenarios for the summaries: a) a real-time scenario where relevant/important messages are pushed to evaluators shortly after their identification and b) an offline schenario, where summaries are produced and sent in batches as email digests. Although the topics of the track are not restricted only to real world events, the task is closely related to the event-based summarization problem, and the approaches presented in section \ref{sec:retrieval_summarization_methods} can be used and evaluated in that setting. The evaluation of participants is based on the methodology presented in \cite{wang2015assessor}.

\section{Conclusions}
\label{sec:conclusions}
The chapter presented an overview of numerous methods for detecting events in large collections and streams of social media content and of methods for retrieving and summarizing content relevant to given events. The proliferation of research on the topic demonstrates the challenging nature of the problem and its potential for creating attractive and useful retrieval applications for end users. 

Throughout the chapter, a number of key important methodological elements were identified that may be considered as components of the presented approaches:
\begin{itemize}
\item \textit{Feature representation:} Text is the primary modality used for representing social media posts, typically using established representations such as Bag-of-Words, which are often enriched by additional more semantically prominent elements such as named entities, tags, and topics. Visual content, when used, has extensively relied on Bag-of-Visual-Words representation, Vector of Locally Aggregated Descriptors, while the recently proposed Deep Convolutional Neural Network features have become increasingly important. Additional modalities such as time, location and social interactions are another important component of social media content in relation to event detection and retrieval, with very simple and ad hoc features being used.
\item \textit{Semantic similarity:} Another important component of event detection methods is the way that the similarity between content items is computed. Standard measures based on cosine and Jaccard similarity are employed by the majority of methods, especially in relation to Bag-of-Words representations, while more sophisticated methods rely on learned similarity models, i.e. models that are trained with indicative pairs of examples to produce a similarity score. In such cases, the training process is crucial in ending up with an accurate and robust similarity function.
\item \textit{Classification models:}  These constitute an essential component in several cases, either as a means of filtering non-event or low-quality content or for classifying media items or clusters of items into distinct categories of events. Most approaches rely on standard models, including Naive Bayes, Logistic Regression, Random Forests and Support Vector Machines, all of which constitute well-tested and easily deployable solutions.
\item \textit{Clustering algorithms:} A crucial component of event detection methods is the clustering algorithm that is used for assigning social media items to clusters corresponding to distinct events. Depending on the input representation, conventional data clustering algorithms ($k$-means, DBSCAN, etc.) or graph-based ones (Spectral, SCAN, QCA, etc.) can be used. Additionally, an important decision is whether the algorithm should lead to a partition of the collection (i.e. all items assigned to clusters) or a selective clustering (as in the case of DBSCAN and SCAN).
\item \textit{Content ranking:} This is a key element of event retrieval and summarization methods and entails a variety of approaches for predicting a score that expresses how suitable a given social media item is as a result to an event query. Important desiderata in this context are the capturing of both relevance and diversity in a selected set of items. Relatively under-explored areas include the personalization of summaries and the consideration of additional criteria such as credibility (especially in the context of news), aesthetics/quality and viewpoint.
\end{itemize}
For researchers and practitioners in the area, who use existing methods or develop new ones, it is always useful to think of the aforementioned elements as design choices that could affect the performance of the resulting system.

An additional finding of our research is the relative scarcity of public resources for evaluating the performance of event detection, retrieval and summarization methods. This is primarily due to the fact that considerable amounts of manual effort are needed to create ground truth annotations for large-scale datasets. Another complication in generating persistent datasets is the highly volatile nature of social media content, sizable portions of which tend to disappear in the long run. Finally, there is a need for cross-platform datasets referring to the same real-world events. In that way, it will be possible to study interesting research questions, such as whether different social media platforms tend to be used in relation to different aspects of an event. To sum up, enriching existing data resources, generating additional ones and conducting comprehensive comparative studies among the multitude of methods that have been proposed would all be valuable contributions towards achieving progress in the field.


\bibliography{bibtex}

\begin{thebibliography}{10}

\bibitem{aggarwal2012event}
Charu~C Aggarwal and Karthik Subbian.
\newblock Event detection in social streams.
\newblock In {\em Proceedings of the 2012 SIAM international conference on data
  mining}, pages 624--635. SIAM, 2012.

\bibitem{allan2002introduction}
James Allan.
\newblock Introduction to topic detection and tracking.
\newblock {\em Topic detection and tracking}, pages 1--16, 2002.

\bibitem{alonso2013timelines}
Omar Alonso and Kyle Shiells.
\newblock Timelines as summaries of popular scheduled events.
\newblock In {\em Proceedings of the 22nd International Conference on World
  Wide Web}, pages 1037--1044. ACM, 2013.

\bibitem{alvanaki2012see}
Foteini Alvanaki, Sebastian Michel, Krithi Ramamritham, and Gerhard Weikum.
\newblock See what's enblogue: real-time emergent topic identification in
  social media.
\newblock In {\em Proceedings of the 15th International Conference on Extending
  Database Technology}, pages 336--347. ACM, 2012.

\bibitem{amati2011fub}
Gianni Amati, Giuseppe Amodeo, Marco Bianchi, Giuseppe Marcone, Fondazione~Ugo
  Bordoni, Carlo Gaibisso, Giorgio Gambosi, Alessandro Celi, Cesidio Di~Nicola,
  and Michele Flammini.
\newblock Fub, iasi-cnr, univaq at trec 2011 microblog track.
\newblock In {\em TREC}, 2011.

\bibitem{amigo2009comparison}
Enrique Amig{\'o}, Julio Gonzalo, Javier Artiles, and Felisa Verdejo.
\newblock A comparison of extrinsic clustering evaluation metrics based on
  formal constraints.
\newblock {\em Information retrieval}, 12(4):461--486, 2009.

\bibitem{bao2013social}
Bing-Kun Bao, Weiqing Min, Ke~Lu, and Changsheng Xu.
\newblock Social event detection with robust high-order co-clustering.
\newblock In {\em Proceedings of the 3rd ACM conference on International
  conference on multimedia retrieval}, pages 135--142. ACM, 2013.

\bibitem{becker2011automatic}
Hila Becker, Feiyang Chen, Dan Iter, Mor Naaman, and Luis Gravano.
\newblock Automatic identification and presentation of twitter content for
  planned events.
\newblock In {\em ICWSM}, 2011.

\bibitem{becker2012identifying}
Hila Becker, Dan Iter, Mor Naaman, and Luis Gravano.
\newblock Identifying content for planned events across social media sites.
\newblock In {\em Proceedings of the fifth ACM international conference on Web
  search and data mining}, pages 533--542. ACM, 2012.

\bibitem{becker2009event}
Hila Becker, Mor Naaman, and Luis Gravano.
\newblock Event identification in social media.
\newblock In {\em WebDB}, 2009.

\bibitem{becker2010learning}
Hila Becker, Mor Naaman, and Luis Gravano.
\newblock Learning similarity metrics for event identification in social media.
\newblock In {\em Proceedings of the third ACM international conference on Web
  search and data mining}, pages 291--300. ACM, 2010.

\bibitem{becker2011beyond}
Hila Becker, Mor Naaman, and Luis Gravano.
\newblock Beyond trending topics: Real-world event identification on twitter.
\newblock {\em ICWSM}, 11(2011):438--441, 2011.

\bibitem{benson2011event}
Edward Benson, Aria Haghighi, and Regina Barzilay.
\newblock Event discovery in social media feeds.
\newblock In {\em Proceedings of the 49th Annual Meeting of the Association for
  Computational Linguistics: Human Language Technologies-Volume 1}, pages
  389--398. Association for Computational Linguistics, 2011.

\bibitem{bian2013multimedia}
Jingwen Bian, Yang Yang, and Tat-Seng Chua.
\newblock Multimedia summarization for trending topics in microblogs.
\newblock In {\em Proceedings of the 22nd ACM international conference on
  Conference on information \& knowledge management}, pages 1807--1812. ACM,
  2013.

\bibitem{blei2002latent}
David~M Blei, Andrew~Y Ng, and Michael~I Jordan.
\newblock Latent dirichlet allocation.
\newblock In {\em Advances in neural information processing systems}, pages
  601--608, 2002.

\bibitem{brenner2013mediaeval}
Markus Brenner and Ebroul Izquierdo.
\newblock Mediaeval 2013: Social event detection, retrieval and classification
  in collaborative photo collections.
\newblock {\em MediaEval}, 1043, 2013.

\bibitem{cai2015popular}
Hongyun Cai, Yang Yang, Xuefei Li, and Zi~Huang.
\newblock What are popular: exploring twitter features for event detection,
  tracking and visualization.
\newblock In {\em Proceedings of the 23rd ACM international conference on
  Multimedia}, pages 89--98. ACM, 2015.

\bibitem{cataldi2010emerging}
Mario Cataldi, Luigi Di~Caro, and Claudio Schifanella.
\newblock Emerging topic detection on twitter based on temporal and social
  terms evaluation.
\newblock In {\em Proceedings of the tenth international workshop on multimedia
  data mining}, page~4. ACM, 2010.

\bibitem{chakrabarti2011event}
Deepayan Chakrabarti and Kunal Punera.
\newblock Event summarization using tweets.
\newblock {\em ICWSM}, 11:66--73, 2011.

\bibitem{chen2009event}
Ling Chen and Abhishek Roy.
\newblock Event detection from flickr data through wavelet-based spatial
  analysis.
\newblock In {\em Proceedings of the 18th ACM conference on Information and
  knowledge management}, pages 523--532. ACM, 2009.

\bibitem{choi2012temporal}
Jaeho Choi and W~Bruce Croft.
\newblock Temporal models for microblogs.
\newblock In {\em Proceedings of the 21st ACM international conference on
  Information and knowledge management}, pages 2491--2494. ACM, 2012.

\bibitem{choi2012quality}
Jaeho Choi, W~Bruce Croft, and Jin~Young Kim.
\newblock Quality models for microblog retrieval.
\newblock In {\em Proceedings of the 21st ACM international conference on
  Information and knowledge management}, pages 1834--1838. ACM, 2012.

\bibitem{chua2013automatic}
Freddy Chong~Tat Chua and Sitaram Asur.
\newblock Automatic summarization of events from social media.
\newblock In {\em ICWSM}, 2013.

\bibitem{denman2014saivt}
Simon Denman, David Dean, Clinton Fookes, and Sridha Sridharan.
\newblock Saivt-admrg@ mediaeval 2014 social event detection.
\newblock In {\em Working Notes Proceedings of the MediaEval 2014 Multimedia
  Benchmark Workshop}, volume 1263, pages 1--2. CEUR Workshop Proceedings,
  2014.

\bibitem{diao2013unified}
Qiming Diao and Jing Jiang.
\newblock A unified model for topics, events and users on twitter.
\newblock ACL, 2013.

\bibitem{dork2010visual}
Marian Dork, Daniel Gruen, Carey Williamson, and Sheelagh Carpendale.
\newblock A visual backchannel for large-scale events.
\newblock {\em IEEE transactions on visualization and computer graphics},
  16(6):1129--1138, 2010.

\bibitem{erkan2004lexrank}
G{\"u}nes Erkan and Dragomir~R Radev.
\newblock Lexrank: Graph-based lexical centrality as salience in text
  summarization.
\newblock {\em Journal of Artificial Intelligence Research}, 22:457--479, 2004.

\bibitem{erosheva2004mixed}
Elena Erosheva, Stephen Fienberg, and John Lafferty.
\newblock Mixed-membership models of scientific publications.
\newblock {\em Proceedings of the National Academy of Sciences}, 101(suppl
  1):5220--5227, 2004.

\bibitem{fung2005parameter}
Gabriel Pui~Cheong Fung, Jeffrey~Xu Yu, Philip~S Yu, and Hongjun Lu.
\newblock Parameter free bursty events detection in text streams.
\newblock In {\em Proceedings of the 31st international conference on Very
  large data bases}, pages 181--192. VLDB Endowment, 2005.

\bibitem{gao2012joint}
Wei Gao, Peng Li, and Kareem Darwish.
\newblock Joint topic modeling for event summarization across news and social
  media streams.
\newblock In {\em Proceedings of the 21st ACM international conference on
  Information and knowledge management}, pages 1173--1182. ACM, 2012.

\bibitem{guille2014mention}
Adrien Guille and C{\'e}cile Favre.
\newblock Mention-anomaly-based event detection and tracking in twitter.
\newblock In {\em Advances in Social Networks Analysis and Mining (ASONAM),
  2014 IEEE/ACM International Conference on}, pages 375--382. IEEE, 2014.

\bibitem{guinaudeau2014limsi}
Camille Guinaudeau, Antoine Laurent, and Herv{\'e} Bredin.
\newblock Limsi@ mediaeval sed 2014.

\bibitem{he2007analyzing}
Qi~He, Kuiyu Chang, and Ee-Peng Lim.
\newblock Analyzing feature trajectories for event detection.
\newblock In {\em Proceedings of the 30th annual international ACM SIGIR
  conference on Research and development in information retrieval}, pages
  207--214. ACM, 2007.

\bibitem{hoang2012ugent}
Thong Hoang Van~Duc, Thomas Demeester, Johannes Deleu, Piet Demeester, and
  Chris Develder.
\newblock Ugent participation in the microblog track 2012.
\newblock In {\em Text Retrieval Conference (TREC-2012)}, pages 1--5, 2012.

\bibitem{hu2012lda}
Yuheng Hu, Ajita John, Fei Wang, and Subbarao Kambhampati.
\newblock Et-lda: Joint topic modeling for aligning events and their twitter
  feedback.
\newblock In {\em AAAI}, volume~12, pages 59--65, 2012.

\bibitem{itikavit}
Kshitij~Gautam Itika~Gupta and Krishna Chandramouli.
\newblock Vit@ mediaeval 2013 social event detection task: Semantic structuring
  of complementary information for clustering events.
\newblock 2013.

\bibitem{lee2012mining}
Chung-Hong Lee.
\newblock Mining spatio-temporal information on microblogging streams using a
  density-based online clustering method.
\newblock {\em Expert Systems with Applications}, 39(10):9623--9641, 2012.

\bibitem{li2012twevent}
Chenliang Li, Aixin Sun, and Anwitaman Datta.
\newblock Twevent: segment-based event detection from tweets.
\newblock In {\em Proceedings of the 21st ACM international conference on
  Information and knowledge management}, pages 155--164. ACM, 2012.

\bibitem{lin2012generating}
Chen Lin, Chun Lin, Jingxuan Li, Dingding Wang, Yang Chen, and Tao Li.
\newblock Generating event storylines from microblogs.
\newblock In {\em Proceedings of the 21st ACM international conference on
  Information and knowledge management}, pages 175--184. ACM, 2012.

\bibitem{lin2003automatic}
Chin-Yew Lin and Eduard Hovy.
\newblock Automatic evaluation of summaries using n-gram co-occurrence
  statistics.
\newblock In {\em Proceedings of the 2003 Conference of the North American
  Chapter of the Association for Computational Linguistics on Human Language
  Technology-Volume 1}, pages 71--78. Association for Computational
  Linguistics, 2003.

\bibitem{liu2011finding}
Xueliang Liu, Rapha{\"e}l Troncy, and Benoit Huet.
\newblock Finding media illustrating events.
\newblock In {\em Proceedings of the 1st ACM International Conference on
  Multimedia Retrieval}, page~58. ACM, 2011.

\bibitem{manchon2013upc}
Daniel Manchon~Vizuete and Xavier Gir{\'o}~Nieto.
\newblock Upc at mediaeval 2013 social event detection task.
\newblock In {\em Proceedings of the MediaEval 2013 Multimedia Benchmark
  Workshop}. CEUR Workshop Proceedings, 2013.

\bibitem{manchon2014upc}
Daniel Manchon-Vizuete, Irene Gris-Sarabia, and Xavier Gir{\'o}~Nieto.
\newblock Upc at mediaeval 2014 social event detection task.
\newblock In {\em MediaEval}, 2014.

\bibitem{marcus2011twitinfo}
Adam Marcus, Michael~S Bernstein, Osama Badar, David~R Karger, Samuel Madden,
  and Robert~C Miller.
\newblock Twitinfo: aggregating and visualizing microblogs for event
  exploration.
\newblock In {\em Proceedings of the SIGCHI conference on Human factors in
  computing systems}, pages 227--236. ACM, 2011.

\bibitem{mathioudakis2010twittermonitor}
Michael Mathioudakis and Nick Koudas.
\newblock Twittermonitor: trend detection over the twitter stream.
\newblock In {\em Proceedings of the 2010 ACM SIGMOD International Conference
  on Management of data}, pages 1155--1158. ACM, 2010.

\bibitem{mcminn2013building}
Andrew~J McMinn, Yashar Moshfeghi, and Joemon~M Jose.
\newblock Building a large-scale corpus for evaluating event detection on
  twitter.
\newblock In {\em Proceedings of the 22nd ACM international conference on
  Information \& Knowledge Management}, pages 409--418. ACM, 2013.

\bibitem{mcparlane2014picture}
Philip~J McParlane, Andrew~James McMinn, and Joemon~M Jose.
\newblock Picture the scene...;: Visually summarising social media events.
\newblock In {\em Proceedings of the 23rd ACM International Conference on
  Conference on Information and Knowledge Management}, pages 1459--1468. ACM,
  2014.

\bibitem{meladianos2015degeneracy}
Polykarpos Meladianos, Giannis Nikolentzos, Fran{\c{c}}ois Rousseau, Yannis
  Stavrakas, and Michalis Vazirgiannis.
\newblock Degeneracy-based real-time sub-event detection in twitter stream.
\newblock In {\em ICWSM}, pages 248--257, 2015.

\bibitem{metzler2012structured}
Donald Metzler, Congxing Cai, and Eduard Hovy.
\newblock Structured event retrieval over microblog archives.
\newblock In {\em Proceedings of the 2012 Conference of the North American
  Chapter of the Association for Computational Linguistics: Human Language
  Technologies}, pages 646--655. Association for Computational Linguistics,
  2012.

\bibitem{miyanishi2013improving}
Taiki Miyanishi, Kazuhiro Seki, and Kuniaki Uehara.
\newblock Improving pseudo-relevance feedback via tweet selection.
\newblock In {\em Proceedings of the 22nd ACM international conference on
  Information \& Knowledge Management}, pages 439--448. ACM, 2013.

\bibitem{moran2016enhancing}
Sean Moran, Richard McCreadie, Craig Macdonald, and Iadh Ounis.
\newblock Enhancing first story detection using word embeddings.
\newblock In {\em Proceedings of the 39th International ACM SIGIR conference on
  Research and Development in Information Retrieval}, pages 821--824. ACM,
  2016.

\bibitem{nguyen2013event}
Truc-Vien Nguyen, Minh-Son Dao, Riccardo Mattivi, Emanuele Sansone,
  Francesco~GB De~Natale, and Giulia Boato.
\newblock Event clustering and classification from social media:
  Watershed-based and kernel methods.
\newblock In {\em MediaEval}, 2013.

\bibitem{nichols2012summarizing}
Jeffrey Nichols, Jalal Mahmud, and Clemens Drews.
\newblock Summarizing sporting events using twitter.
\newblock In {\em Proceedings of the 2012 ACM international conference on
  Intelligent User Interfaces}, pages 189--198. ACM, 2012.

\bibitem{papadopoulos2012social}
Symeon Papadopoulos, Emmanouil Schinas, Vasileios Mezaris, Raphaël Troncy, and
  Yiannis Kompatsiaris.
\newblock Social event detection at mediaeval 2012: Challenges, dataset and
  evaluation.
\newblock In {\em MediaEval}, 2012.

\bibitem{papadopoulos2011social}
Symeon Papadopoulos, Raphael Troncy, Vasileios Mezaris, Benoit Huet, and
  Ioannis Kompatsiaris.
\newblock Social event detection at mediaeval 2011: Challenges, dataset and
  evaluation.
\newblock In {\em MediaEval}, 2011.

\bibitem{parikh2013events}
Ruchi Parikh and Kamalakar Karlapalem.
\newblock Et: events from tweets.
\newblock In {\em Proceedings of the 22nd international conference on world
  wide web}, pages 613--620. ACM, 2013.

\bibitem{paul2009cross}
Michael Paul and Roxana Girju.
\newblock Cross-cultural analysis of blogs and forums with mixed-collection
  topic models.
\newblock In {\em Proceedings of the 2009 Conference on Empirical Methods in
  Natural Language Processing: Volume 3-Volume 3}, pages 1408--1417.
  Association for Computational Linguistics, 2009.

\bibitem{paul2010two}
Michael Paul and Roxana Girju.
\newblock A two-dimensional topic-aspect model for discovering multi-faceted
  topics.
\newblock {\em Urbana}, 51(61801):36, 2010.

\bibitem{petkos2012social}
Georgios Petkos, Symeon Papadopoulos, and Yiannis Kompatsiaris.
\newblock Social event detection using multimodal clustering and integrating
  supervisory signals.
\newblock In {\em Proceedings of the 2nd ACM International Conference on
  Multimedia Retrieval}, page~23. ACM, 2012.

\bibitem{petkos2014social}
Georgios Petkos, Symeon Papadopoulos, Vasileios Mezaris, and Yiannis
  Kompatsiaris.
\newblock Social event detection at mediaeval 2014: Challenges, datasets, and
  evaluation.
\newblock In {\em MediaEval}, 2014.

\bibitem{petkos_sewm2014}
Georgios Petkos, Symeon Papadopoulos, Vasileios Mezaris, Raphael Troncy,
  Philipp Cimiano, Timo Reuter, and Yiannis Kompatsiaris.
\newblock Social event detection at mediaeval: a three-year retrospect of tasks
  and results.
\newblock In {\em Proceedings of International Workshop on Social Events in Web
  Multimedia (co-located with ICMR 2014)}, pages 27--34, 2014.

\bibitem{petkos_mtap2017}
Georgios Petkos, Manos Schinas, Symeon Papadopoulos, and Yiannis Kompatsiaris.
\newblock Graph-based multimodal clustering for social multimedia.
\newblock {\em Multimedia Tools and Applications}, 76(6):7897--–7919, 2017.

\bibitem{petrovic2010streaming}
Sa{\v{s}}a Petrovi{\'c}, Miles Osborne, and Victor Lavrenko.
\newblock Streaming first story detection with application to twitter.
\newblock In {\em Human Language Technologies: The 2010 Annual Conference of
  the North American Chapter of the Association for Computational Linguistics},
  pages 181--189. Association for Computational Linguistics, 2010.

\bibitem{petrovic2012using}
Sa{\v{s}}a Petrovi{\'c}, Miles Osborne, and Victor Lavrenko.
\newblock Using paraphrases for improving first story detection in news and
  twitter.
\newblock In {\em Proceedings of the 2012 Conference of the North American
  Chapter of the Association for Computational Linguistics: Human Language
  Technologies}, pages 338--346. Association for Computational Linguistics,
  2012.

\bibitem{rafailidis2013data}
Dimitrios Rafailidis, Theodoros Semertzidis, Michalis Lazaridis, Michael~G
  Strintzis, and Petros Daras.
\newblock A data-driven approach for social event detection.
\newblock In {\em MediaEval}, 2013.

\bibitem{reuter2012event}
Timo Reuter and Philipp Cimiano.
\newblock Event-based classification of social media streams.
\newblock In {\em Proceedings of the 2nd ACM International Conference on
  Multimedia Retrieval}, page~22. ACM, 2012.

\bibitem{reuter2013social}
Timo Reuter, Symeon Papadopoulos, Giorgos Petkos, Vasileios Mezaris, Yiannis
  Kompatsiaris, Philipp Cimiano, Christopher de~Vries, and Shlomo Geva.
\newblock Social event detection at mediaeval 2013: Challenges, datasets, and
  evaluation.
\newblock In {\em Proceedings of the MediaEval 2013 Multimedia Benchmark
  Workshop Barcelona, Spain, October 18-19, 2013}, 2013.

\bibitem{riga2014certh}
Marina Riga, Georgios Petkos, Symeon Papadopoulos, Manos Schinas, and Yiannis
  Kompatsiaris.
\newblock Certh@ mediaeval 2014 social event detection task.
\newblock In {\em MediaEval}, 2014.

\bibitem{ritter2012open}
Alan Ritter, Oren Etzioni, Sam Clark, et~al.
\newblock Open domain event extraction from twitter.
\newblock In {\em Proceedings of the 18th ACM SIGKDD international conference
  on Knowledge discovery and data mining}, pages 1104--1112. ACM, 2012.

\bibitem{sakaki_www2010}
Takeshi Sakaki, Makoto Okazaki, and Yutaka Matsuo.
\newblock Earthquake shakes twitter users: real-time event detection by social
  sensors.
\newblock In {\em Proceedings of the 19th international conference on World
  wide web}, pages 851--860. ACM, 2010.

\bibitem{samangooei2013social}
Sina Samangooei, Jonathon Hare, David Dupplaw, Mahesan Niranjan, Nicholas
  Gibbins, Paul~H Lewis, Jamie Davies, Neha Jain, and John Preston.
\newblock Social event detection via sparse multi-modal feature selection and
  incremental density based clustering.
\newblock 2013.

\bibitem{sankaranarayanan2009twitterstand}
Jagan Sankaranarayanan, Hanan Samet, Benjamin~E Teitler, Michael~D Lieberman,
  and Jon Sperling.
\newblock Twitterstand: news in tweets.
\newblock In {\em Proceedings of the 17th acm sigspatial international
  conference on advances in geographic information systems}, pages 42--51. ACM,
  2009.

\bibitem{sayyadi2009event}
Hassan Sayyadi, Matthew Hurst, and Alexey Maykov.
\newblock Event detection and tracking in social streams.
\newblock In {\em Icwsm}, 2009.

\bibitem{schinas2013mediaeval}
Emmanouil Schinas, Eleni Mantziou, Symeon Papadopoulos, Georgios Petkos, and
  Yiannis Kompatsiaris.
\newblock Certh @ mediaeval 2013 social event detection task.
\newblock {\em MediaEval}, 1043, 2013.

\bibitem{schinas_icmr2015}
Manos Schinas, Symeon Papadopoulos, Yiannis Kompatsiaris, and Pericles~A.
  Mitkas.
\newblock Visual event summarization on social media using topic modelling and
  graph-based ranking algorithms.
\newblock In {\em Proceedings of the 5th ACM on International Conference on
  Multimedia Retrieval}, ICMR '15, pages 203--210, New York, NY, USA, 2015.
  ACM.

\bibitem{schinas_ijmir2016}
Manos Schinas, Symeon Papadopoulos, Yiannis Kompatsiaris, and Pericles~A
  Mitkas.
\newblock Mgraph: multimodal event summarization in social media using topic
  models and graph-based ranking.
\newblock {\em International Journal of Multimedia Information Retrieval},
  5(1):51--69, 2016.

\bibitem{severyn2015learning}
Aliaksei Severyn and Alessandro Moschitti.
\newblock Learning to rank short text pairs with convolutional deep neural
  networks.
\newblock In {\em Proceedings of the 38th International ACM SIGIR Conference on
  Research and Development in Information Retrieval}, pages 373--382. ACM,
  2015.

\bibitem{shaw2009lode}
Ryan Shaw, Rapha{\"e}l Troncy, and Lynda Hardman.
\newblock Lode: Linking open descriptions of events.
\newblock {\em ASWC}, 9:153--167, 2009.

\bibitem{shen2013participant}
Chao Shen, Fei Liu, Fuliang Weng, and Tao Li.
\newblock A participant-based approach for event summarization using twitter
  streams.
\newblock In {\em HLT-NAACL}, pages 1152--1162, 2013.

\bibitem{strehl2002cluster}
Alexander Strehl and Joydeep Ghosh.
\newblock Cluster ensembles-a knowledge reuse framework for combining
  partitionings.
\newblock In {\em Aaai/iaai}, pages 93--99, 2002.

\bibitem{sutanto2013admrg}
Taufik Sutanto and Richi Nayak.
\newblock Admrg@ mediaeval 2013 social event detection.
\newblock In {\em Proceedings of the MediaEval 2013 Multimedia Benchmark
  Workshop}, volume 1043. CEUR-WS. org, 2013.

\bibitem{sutanto2014ranking}
Taufik Sutanto and Richi Nayak.
\newblock Ranking based clustering for social event detection.
\newblock In {\em MediaEval}, 2014.

\bibitem{thomee2016yfcc100m}
Bart Thomee, David~A Shamma, Gerald Friedland, Benjamin Elizalde, Karl Ni,
  Douglas Poland, Damian Borth, and Li-Jia Li.
\newblock Yfcc100m: The new data in multimedia research.
\newblock {\em Communications of the ACM}, 59(2):64--73, 2016.

\bibitem{tserpessimilarity}
Konstantinos Tserpes, Magdalini Kardara, and Theodora Varvarigou.
\newblock A similarity-based chinese restaurant process for social event
  detection.

\bibitem{wang2012generating}
Dingding Wang, Tao Li, and Mitsunori Ogihara.
\newblock Generating pictorial storylines via minimum-weight connected
  dominating set approximation in multi-view graphs.
\newblock In {\em AAAI}, 2012.

\bibitem{wang2006topics}
Xuerui Wang and Andrew McCallum.
\newblock Topics over time: a non-markov continuous-time model of topical
  trends.
\newblock In {\em Proceedings of the 12th ACM SIGKDD international conference
  on Knowledge discovery and data mining}, pages 424--433. ACM, 2006.

\bibitem{wang2012social}
Yanxiang Wang, Hari Sundaram, and Lexing Xie.
\newblock Social event detection with interaction graph modeling.
\newblock In {\em Proceedings of the 20th ACM international conference on
  Multimedia}, pages 865--868. ACM, 2012.

\bibitem{wang2015assessor}
Yulu Wang, Garrick Sherman, Jimmy Lin, and Miles Efron.
\newblock Assessor differences and user preferences in tweet timeline
  generation.
\newblock In {\em Proceedings of the 38th International ACM SIGIR Conference on
  Research and Development in Information Retrieval}, pages 615--624. ACM,
  2015.

\bibitem{wei2015bayesian}
Wei Wei, Kenneth Joseph, Wei Lo, and Kathleen~M Carley.
\newblock A bayesian graphical model to discover latent events from twitter.
\newblock In {\em ICWSM}, pages 503--512, 2015.

\bibitem{weng2011event}
Jianshu Weng and Bu-Sung Lee.
\newblock Event detection in twitter.
\newblock {\em ICWSM}, 11:401--408, 2011.

\bibitem{wistuba2013supervised}
Martin Wistuba and Lars Schmidt-Thieme.
\newblock Supervised clustering of social media streams.
\newblock In {\em MediaEval}, 2013.

\bibitem{you2013geam}
Yue You, Guangyan Huang, Jian Cao, Enhong Chen, Jing He, Yanchun Zhang, and
  Liang Hu.
\newblock Geam: A general and event-related aspects model for twitter event
  detection.
\newblock In {\em International Conference on Web Information Systems
  Engineering}, pages 319--332. Springer, Berlin, Heidelberg, 2013.

\bibitem{zaharieva2014clustering}
Maia Zaharieva, Daniel Schopfhauser, Manfred Del~Fabro, and Matthias
  Zeppelzauer.
\newblock Clustering and retrieval of social events in flickr.
\newblock In {\em MediaEval}, 2014.

\bibitem{zeppelzauer2013unsupervised}
Matthias Zeppelzauer, Maia Zaharieva, and Manfred Del~Fabro.
\newblock Unsupervised clustering of social events.
\newblock In {\em MediaEval}, 2013.

\bibitem{zhang2015event}
Xiaoming Zhang, Xiaoming Chen, Yan Chen, Senzhang Wang, Zhoujun Li, and Jiali
  Xia.
\newblock Event detection and popularity prediction in microblogging.
\newblock {\em Neurocomputing}, 149:1469--1480, 2015.

\bibitem{zhao2011comparing}
Wayne~Xin Zhao, Jing Jiang, Jianshu Weng, Jing He, Ee-Peng Lim, Hongfei Yan,
  and Xiaoming Li.
\newblock Comparing twitter and traditional media using topic models.
\newblock In {\em European Conference on Information Retrieval}, pages
  338--349. Springer, 2011.

\bibitem{zhou2015unsupervised}
Deyu Zhou, Liangyu Chen, and Yulan He.
\newblock An unsupervised framework of exploring events on twitter: Filtering,
  extraction and categorization.
\newblock In {\em AAAI}, pages 2468--2475, 2015.

\bibitem{zhou2014event}
Xiangmin Zhou and Lei Chen.
\newblock Event detection over twitter social media streams.
\newblock {\em The VLDB journal}, 23(3):381--400, 2014.

\end{thebibliography}
\end{document}